\documentclass[a4paper,twocolumn,11pt,accepted=2017-05-09]{quantumarticle}
\pdfoutput=1
\usepackage[utf8]{inputenc}
\usepackage[english]{babel}
\usepackage[T1]{fontenc}
\usepackage{amsmath}
\usepackage{hyperref}
\usepackage{amssymb}
\usepackage{float}

\usepackage{tikz}
\usepackage{lipsum}

\begin{document}

\title{Quantum Secret Sharing Enhanced: Utilizing W States for Anonymous and Secure Communication}

\author{Guo-Dong Li}
\affiliation{School of Control and Computer Engineering, North China Electric Power University, Beijing 102206, China}
\author{Wen-Chuan Cheng}
\affiliation{School of Control and Computer Engineering, North China Electric Power University, Beijing 102206, China}
\author{Qing-Le Wang}
\email{wqle519@gmail.com}
\affiliation{School of Control and Computer Engineering, North China Electric Power University, Beijing 102206, China}
\affiliation{Key Lab of Information Network Security, Ministry of Public Security, Shanghai 200031, China}
\author{Long Cheng}
\affiliation{School of Control and Computer Engineering, North China Electric Power University, Beijing 102206, China}
\author{Ying Mao}
\affiliation{Department of Computer and Information Science, Fordham University, New York City 10458, USA}
\author{Heng-Yue Jia}
\affiliation{School of Information, Central University of Finance and Economics, Beijing 102206, China}
\affiliation{Engineering Research Center of State Financial Security, Ministry of Education, Central University of Finance and Economics, Beijing 102206, China}
\maketitle

\begin{abstract}
  Quantum secret sharing (QSS) is the result of merging the principles of quantum mechanics with secret information sharing. It enables a sender to share a secret among receivers, and the receivers can then collectively recover the secret when the need arises. To enhance the practicality of these quantum protocols, an innovative concept of quantum anonymous secret sharing (QASS) is advanced. In this paper, we propose a QASS protocol via W states, which can share secrets while ensuring recover-ability, recover-security, and recover-anonymity. We have rigorously evaluated our protocols, verifying their accuracy and fortifying their security against scenarios involving the active adversary. This includes considerations for dishonest receivers and non-receivers. Moreover, acknowledging the imperfections inherent in real-world communication channels, we have also undertaken an exhaustive analysis of our protocol's security and effectiveness in a quantum network where some form of noise is present. Our investigations reveal that W states exhibit good performance in mitigating noise interference, making them apt for practical applications. 
\end{abstract}

\section{Introduction}

Anonymity is an important cryptographic property. With an increasing emphasis on personal privacy by communication users, the anonymity of user identities and the confidentiality of information\cite{ref1,ref2,ref3} should hold equal importance. There are various classical cryptographic applications emphasizing anonymity, such as anonymous voting\cite{ref4,ref5}, anonymous key distribution\cite{ref6,ref7}, and anonymous 
private information retrieval\cite{ref8,ref9}, have been developed. The anonymous secret sharing (ASS) technique, which we discuss, is also applicable across many cryptographic domains, such as secure key management and multiparty secure conferences, and more. However, while classical ASS schemes have substantial practical value\cite{ref10}, they rely on the computational complexity of classical encryption, making them potentially vulnerable to adversaries with strong computational capabilities. Fortunately, QASS addresses this issue while ensuring information-theoretic security. 

In 2023, Li et al. proposed the first authenticated anonymous secret sharing protocol based on $d$-dimensional quantum systems, aiming to address the anonymity issue of receiver identities in the secret sharing process\cite{ref11}. In this protocol, secret senders authenticate participants using GHZ states and construct anonymous entanglement among a specified set of anonymous receivers, ultimately sharing classical information. The constructed anonymous entanglement provides dual protection for secret sharing tasks, ensuring both message confidentiality and receiver identity anonymity.

Anonymous entanglement is the core of QASS, achieved by performing local operations at nodes in the network to create entanglement links between senders and anonymous receivers. It also plays a crucial role in protecting user identity anonymity in quantum cryptography, with many protocols proposed for various tasks such as anonymous ranking\cite{ref12,ref13,ref14}, voting\cite{ref15,ref16,ref17}, and communication\cite{ref18,ref19,ref20}. Among these, GHZ states are the most commonly used anonymous entanglement resource. However, in noisy scenarios, the fidelity of the anonymous entanglement GHZ state may be poor, requiring high channel demands. Victoria Lipinska's research suggests that using W states to construct anonymous entanglement has advantages in terms of operational simplicity, performance in noisy channels, and more\cite{ref21}.

Pioneer QASS proposed by Li et al. accomplished the task of sharing classical information \cite{ref11}. To our knowledge, sharing quantum information is also an important branch of quantum secret sharing, and has the same important status as sharing classical information. Its main advantage is the ability to achieve direct transmission of quantum information, which is particularly important for quantum computing and quantum communication networks. In this paper, leveraging the advantages of W states in constructing anonymous entanglement, we design a novel QASS protocol based on W states capable of accomplishing tasks involving the sharing of classical or quantum information. We elaborate on the process of anonymously sharing secret information and present corresponding sub-protocols. These sub-protocols include quantum identity authentication, quantum notification, anonymous entanglement, anonymous secret sharing, and anonymous secret recovery protocols. Our research demonstrates that W states can be applied to the anonymous secret sharing process. Additionally, we thoroughly prove the security of the protocol in an active adversary scenario, covering message confidentiality, identity information privacy, and participant identity anonymity. We emphasize that the security of our protocol remains unchanged when all particles experience the same type of noise interference. Importantly, we consider the feasibility of anonymous secret sharing in noisy quantum networks, quantifying the performance of our protocol through the fidelity of transmitted quantum states. Besides, anonymous entanglement constructed using W states can tolerate an unresponsive node\cite{ref21}. For example, if one of the qubits of a multi-partite state gets lost.

The rest paper is organized as follows. In Section 2, we define anonymous secret sharing and related concepts. Section 3 proposes the QASS protocol based on W states, including the relevant sub-protocols. In Section 4, we validate the correctness of the protocol, provide security definitions, and prove the protocol's security in active adversary scenarios. In Section 5, we explore the security and performance of the protocol in noisy quantum networks. We draw our conclusion in the last section.

\section{Preliminaries} 

Our $(n, n)$ quantum secret sharing (QSS) scheme necessitates the collaborative effort of all receivers for secret reconstruction. This scheme is specifically designed for high security requirements, incorporating the use of anonymous entanglement to ensure robust protection, especially in applications where recipient anonymity is crucial.

Before presenting a comprehensive definition, it is essential to introduce the concept of an access structure. In a secret sharing protocol, participants can be categorized into three distinct roles: the secret sender, potential receivers, and secret restorer. The qualified subset refers to the group of receivers capable of effectively recovering the secret. Let $[\mathcal{B}] = {Bob_1, Bob_2, \ldots, Bob_n}$ represents the set of potential receivers. A monotone access structure, denoted as $\Gamma \subseteq 2^{\mathcal{B}}$, encompasses all qualified subsets within $[\mathcal{B}]$ \cite{ref10,ref22}. Our $(n, n)$ QASS scheme can be defined in conjunction with its properties based on the access structure as follows:

\textbf{Definition 1:} ($(n, n)$ quantum anonymous secret sharing) \textbf{.} In a $(n, n)$ quantum anonymous secret sharing, $[\mathcal{B}]$ is the set of potential receivers, secret sender chooses $m$ of them to be receivers. Let $[\mathcal{R}] = \{Bobr_1, Bobr_2,$ $ \cdots, Bobr_m\} \subseteq [\mathcal{B}]$ be the set of anonymous receivers, access structure $\Gamma = \{\mathcal{R}\}$. Then we can say a perfect ($n,n$) quantum anonymous secret sharing scheme is a collection of distribution rules that satisfy the following three properties:

$\bullet$ Recover-ability: For a random participant subset $[\mathcal{B}'] \subseteq [\mathcal{B}]$, $[\mathcal{B}'] \in \Gamma$, if all of the participants in $[\mathcal{B}']$ pool their shares, they can determine the value of the secret $K$.

$\bullet$ Recover-security: For a random participant subset $[\mathcal{B}'] \subseteq [\mathcal{B}]$, $[\mathcal{B}'] \notin \Gamma$, then the participants in $[\mathcal{B}']$ can determine nothing about the value of the secret $K$ (in an information-theoretic sense), even with infinite computational resources.

$\bullet$ Receiver-anonymity: The secret sharing and secret recovering processes can guarantee the receiver-anonymous.

In Definition 1, we address the concept of receiver anonymity, a pivotal element within the domain of anonymous communication. Consider an entity, denoted as $Bobr_i$ ($i \in [1,m]$), representing an unidentified anonymous receiver. The network encompasses an adversary whose objective is to ascertain the identity of $Bobr_i$ amidst a pool of potential receivers. This adversary has control over a subset of these potential receivers, which we define as dishonest. Let $[\mathcal{D}] \subseteq [\mathcal{B}]$ represent this set of dishonest potential receivers, and $[\mathcal{H}] \subseteq [\mathcal{B}]$ denote the set of honest potential receivers.

The protocol is deemed receiver-anonymous if the adversary's probability of correctly identifying $Bobr_i$ does not exceed the initial uncertainty regarding $Bobr_i$'s identity before the protocol's initiation. This initial uncertainty is quantified by the prior probability, expressed as $P[br_i = b_j|br_i\notin \mathcal{D}]$. To elucidate, receiver anonymity can be formally defined as follows:

\textbf{Definition 2:} (Receiver-anonymity) \textbf{.} Given that the sender $Alice$ is honest, we say that an anonymous secret sharing protocol is receiver-anonymous if, the probability of the adversary guessing that $Bobr_i$ to be $Bob_j$ is
\begin{align}
    \label{eqn.2}
    & P_{guess}[Bobr_i|\mathcal{W}^{\mathcal{D}}, \mathcal{C}, Bobr_i \notin \mathcal{D}]\nonumber \\
    \leqslant & \mathop{max}\limits_{Bob_j \in [\mathcal{H}]} P[Bobr_i = Bob_j| Bobr_i\notin \mathcal{D}].
\end{align}

Here $\mathcal{W}^{\mathcal{D}}$ denotes the adversary's quantum states distributed by $Alice$, $\mathcal{C}$ denotes all classical and quantum side information accessible to the adversary. In words, the protocol is receiver-anonymous if the probability that the adversary guesses the identity of any anonymous receiver $Bobr_i$ at the end of the protocol is not larger than the probability that an honest $Bob_j$ is a receiver, maximized over all the honest potential receivers.  

Definition 2 articulates anonymity within the context of either a perfect channel model or a noise model where each qubit is uniformly affected by an identical noisy channel. However, this definition requires modification when addressing receiver anonymity under conditions of variable network noise. In practical quantum networks, it is plausible that qubits traversing a noisy channel experience non-uniform noise effects. This study aims to examine scenarios wherein each qubit is subject to marginally distinct noise, a concept we term as $\varepsilon$-receiver security.

\textbf{Definition 3} ($\varepsilon$-receiver anonymity) \textbf{.} Given that the sender $Alice$ is honest, an anonymous secret sharing protocol is $\varepsilon$-receiver-anonymous if, the probability of the adversary guessing that $Bobr_i$ to be $Bob_j$ is
\begin{align}
    &P_{guess}[Bobr_i|\mathcal{W}^{\mathcal{D}}, \mathcal{C}, Bobr_i\notin \mathcal{D}]\nonumber \\
         \leqslant & \mathop{max}\limits_{Bob_j\in \mathcal{H}}P[Bobr_i = Bob_j|Bobr_i\notin \mathcal{D}] + \varepsilon.
\end{align}

Here $\varepsilon$ is a parameter used to characterize the channel noise perturbation. This is to say, if the perturbation is small, the guessing probability in a noisy channel with perturbation is $\varepsilon$-close to the guessing probability in a perfect channel or a noisy channel without perturbation, then it can be said that the protocol is $\varepsilon$-receiver-anonymous.

\section{The Protocol for Anonymous Secret Sharing}

Investigate a quantum network involving three entities: a publicly-known secret sender, denoted as $Alice$; an honest secret restorer, referred to as $Charlie$; and $n$ potential receivers, labeled as $Bob_1, Bob_2, \cdots, Bob_n$. The secret sharing phase involves $Alice$ transmitting a confidential message to a specific subset of receivers within the network, symbolized as $[\mathcal{R}]$. Each selected receiver acquires a fragment of the secret. It is hypothesized that any subset of the potential receivers, including non-receivers and even some receivers, may exhibit corrupt behavior. This corruption could manifest as individual or collective attempts to illicitly access additional secret information or infer the identities of other honest receivers.

$Alice$ functions as the secret sender whose identity is public. $Charlie$, designated as the secret restorer, is responsible for reconstructing the secret during the recovering phase. This reconstruction is based on the measurement outcomes disclosed by $Alice$ and the data shared by other potential receivers. This study excludes the possibility of $Alice$ and $Charlie$ participating in active sabotage or disclosing the identities of others.

Under these premises, a novel quantum anonymous secret sharing protocol is proposed, comprising various sub-protocols.

\subsection{Quantum Identity Authentication Protocol}

To resist impersonation attacks, $Alice$ and each $Bob_j$ ($j\in [1, n]$) use their personal identification number to prepare a single photon token for mutual identity authentication. Before authentication, $Alice$ generates an one-time $PIN^{aj}$, send to $Bob_j$ via QKD\cite{ref23} or a face-to-face way. At the same time, $Bob_j$ generates $PIN^{bj}$ and sends to $Alice$ by the same way. The form of $PIN^{aj}$ and $PIN^{bj}$ are as follows:
\begin{align}
PIN^{aj} &=\{PIN_1^{aj}, PIN_2^{aj}, \cdots, PIN_l^{aj}\}; \nonumber\\
PIN^{bj} &=\{PIN_1^{bj}, PIN_2^{bj}, \cdots, PIN_l^{bj}\};
\end{align}
where $PIN_i^{aj},PIN_i^{bj}\in \{00, 01, 10, 11\}$, $i\in [1, l]$, $j\in [1, n]$, $2l$ is the length of $PIN^{aj}$,$PIN^{bj}$. The authentication process between them is shown in protocol \hyperref[Protocol 1]{1}.

\begin{table}[H]
\label{Protocol 1}
\begin{tabular}{@{}p{1mm}p{2mm}p{72mm}}
\hline
\multicolumn{3}{l}{\textbf{Protocol 1:} identity authentication protocol}\\
\toprule
 & \multicolumn{2}{p{74mm}}{\textbf{Goal:} $Alice$ and $Bob_j$ perform mutual authentication.}\\
 & \multicolumn{2}{p{74mm}}{\textbf{Input:} $Private \{PIN^{aj}_i, PIN^{bj}_i\}_{i=1}^l$ shared by $Alice$ and $Bob_j$ in advance.}\\
  & \multicolumn{2}{l}{\textbf{Output:} Authentication success mark $U_j$.}\\
 & (1) & $Alice$ generates $l$ authentication photons as a token based on $PIN^{aj}$. She then performs operations according to $PIN^{bj}$. The guidelines for photon generation are presented in table \ref{Table 1}, and the operational rules are specified in table \ref{Table 2}.\\
 & (2) & $Alice$ transmits the token to $Bob_j$.\\
 & (3) & Upon receiving the token, $Bob_j$ executes a unitary operation on each photon, following the instructions of $PIN^{bj}$, and employs the measurement basis indicated by $PIN^{aj}$ for each particle.\\
  & (4) & If $Bob_j$'s measurement results aligns with the indication of $PIN^{aj}$, he confirms $Alice$ is trustworthy. Otherwise, he issues an alert regarding the illegitimacy of $Alice$'s identity.\\
\hline
\end{tabular}
\end{table}

\begin{table}[H]
\begin{tabular}{@{}p{1mm}p{2mm}p{72mm}}
\hline
\multicolumn{3}{l}{Continuation of \textbf{Protocol 1}}\\
\hline
 & (5) & If the results conveyed by $Bob_j$ are same to the token prepared by $Alice$, she acknowledges $Bob_j$’s legal identities and sets $U_j=1$. Otherwise, $Alice$ alerts about the illegitimacy of $Bob_j$'s identity and sets $U_j=0$.\\
 \hline
\end{tabular}
\end{table}

\begin{table}[H]
\caption{The use of $PIN^{aj}_i$ in this protocol.}\label{Table 1}
\centering
\begin{tabular}{|c|c|c|c|c|}
\hline
$PIN^{aj}_i$ & 00 & 01 & 10 & 11\\
\hline
$Alice$'s generation & $|0\rangle$ & $|1\rangle$ & $|+\rangle$ & $|-\rangle$\\
\hline
$Bob_j$'s measurement basis & $\mathcal{Z}$ & $\mathcal{Z}$ & $\mathcal{X}$ & $\mathcal{X}$\\
\hline
\end{tabular}
\end{table}

\begin{table}[H]
\caption{The use of $PIN^{bj}_i$ in this protocol.}\label{Table 2}
\centering
\begin{tabular}{|c|c|c|c|c|}
\hline
$PIN^{bj}_i$ & 00 & 01 & 10 & 11\\
\hline
$Alice$'s operation & $I$ & $X$ & $Y$ & $Z$\\
\hline
$Bob_j$'s operation & $I$ & $X$ & $Y$ & $Z$\\
\hline
\end{tabular}
\end{table}

Note that, $\mathcal{Z}$-basis and $\mathcal{X}$-basis are rectilinear basis and diagonal basis commonly used in quantum communication. $I$=$\begin{pmatrix}
    1 & 0\\
    0 & 1
\end{pmatrix}$, $X$=$\begin{pmatrix}
    0 & 1\\
    1 & 0
\end{pmatrix}$, $Y$=$\begin{pmatrix}
    0 & -i\\
    i & 0
\end{pmatrix}$, $Z$=$\begin{pmatrix}
    1 & 0\\
    0 & -1
\end{pmatrix}$.

\subsection{Quantum Anonymous Notification Protocol}

Quantum anonymous notification protocol is designed to solve the problem that $Alice$ secretly informs each anonymous receiver of his identity. $Alice$ will take advantage of protocol \hyperref[Protocol 2]{2} and use $n$ notification states to separately inform each $Bob_i$ whether he is a receiver or not. The process is shown in protocol \hyperref[Protocol 2]{2}.

\begin{table}[H]
\label{Protocol 2}
\begin{tabular}{@{}p{1mm}p{2mm}p{72mm}}
\toprule
\multicolumn{3}{l}{\textbf{Protocol 2:} anonymous notification protocol}\\
\hline
 & \multicolumn{2}{p{75mm}}{\textbf{Goal:} $Alice$ separately notifies $Bob_1, Bob_2,$$ \cdots, Bob_n$ whether he is a receiver or not in an anonymous way.}\\
   & \multicolumn{2}{p{75mm}}{\textbf{Input:} $Alice$'s choice of $m$ receivers.}\\
 & \multicolumn{2}{p{74mm}}{\qquad $Alice$ performs $n$ rounds to notify $Bob_1, Bob_2,$ $ \cdots, Bob_n$. For the $i$-th ($i\in[1,n]$) round:} \\
 \hline
\end{tabular}
\end{table}

\begin{table}[H]
\begin{tabular}{@{}p{1mm}p{2mm}p{72mm}}
\toprule
\multicolumn{3}{l}{Continuation of \textbf{Protocol 2}}\\
\hline
 & (1) & $Alice$ generates a notification W state, whose form is as described in Eq.(\ref{eqn.1}). If $Bob_i$ is a receiver, she selects a random but odd number of particles in this W state and performs the $X$ operator in turn. Otherwise, she selects a random but even number of particles in this W state and performs the $X$ operator in turn. \\
 & (2) & $Alice$ separately sends the first particle, the second particle, $\cdots$, and the $n$-th particle to $Bob_1, Bob_2, \cdots, Bob_n$, and keeps the $(n+1)$-th particle. Then she measures her qubit in the $\mathcal{Z}$-basis. The measured result is denoted as $N^{n+1}_i$. \\
  & (3) & For each participant $Bob_j$ ($j\neq i$), he measures his qubit in the $\mathcal{Z}$-basis, and publishes his measured result, denoted as $N_i^j$. $Bob_i$ also measures and records, but not publishes.\\
 & (4) &  $Bob_i$ calculates $N_i=\oplus_{j=1}^{n+1}N_i^j$. If $N_i$=0, then he identifies himself as a secret receiver. If $N_i$=1, then he identifies himself as a non-receiver.\\
\hline
\end{tabular}
\end{table}

Note that, the $(n+1)$-particle W state used in this protocol is as follows:
\begin{small}
    \begin{align}
\label{eqn.1}
&|W^{n+1}\rangle=\frac{1}{\sqrt{n+m}}(\sqrt{m}|0_10_2\cdots 0_n1_{n+1}\rangle + \nonumber \\
&\qquad |0_10_2\cdots 1_n0_{n+1}\rangle + \cdots + |1_10_2\cdots 0_n0_{n+1}\rangle ).
\end{align}
\end{small}
For convenience, let $W_{i,j}$ be the $j$-th particle in the $i$-th $|W^{n+1}\rangle$.

\subsection{Anonymous Entanglement Protocol}

In the proposed system, the key to sharing secrets anonymously without being known by the adversary is the establishment of an anonymous entanglement state between the sender and the intended receivers. This approach in protocol~\hyperref[Protocol 3]{3} differs significantly from the methods of anonymously generating Bell states or GHZ states as discussed in Ref.~\cite{ref24, ref25}, it is a novel scheme for anonymously establishing W states. Here, $Alice$ prepares and shares $|W^{n+1}\rangle$ with potential receivers in advance and wants to anonymously construct an entanglement state $|\overline{W}^{m+1}\rangle$ with secret receivers.

\begin{figure*}[t]
	\centering
		\includegraphics[width=1.0\linewidth]{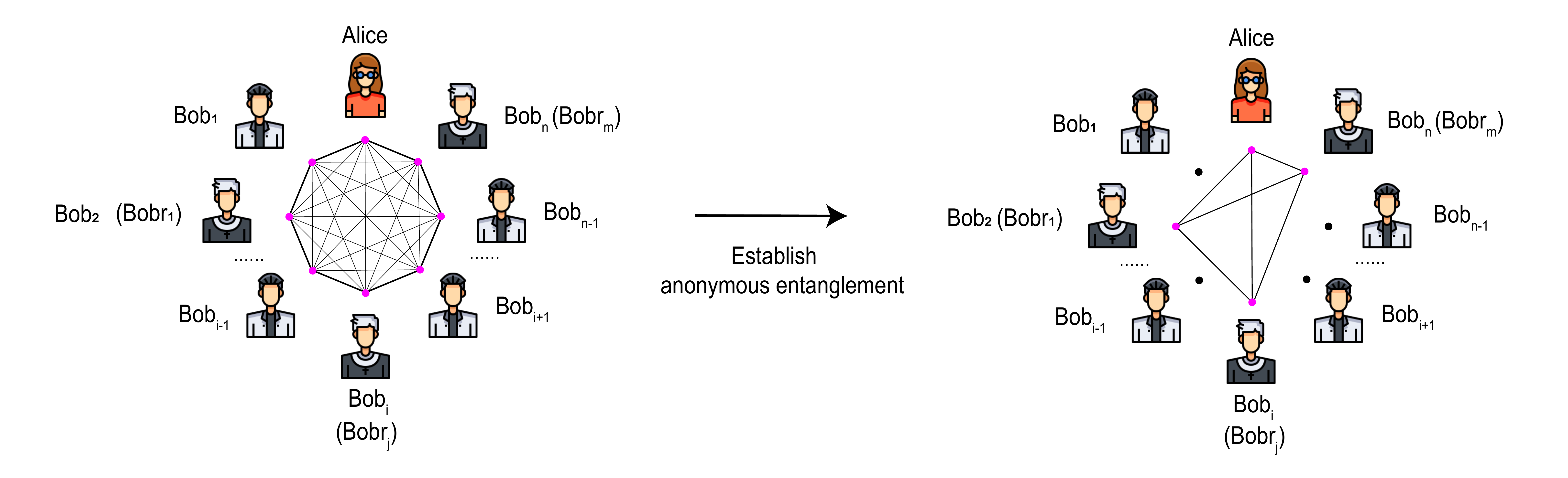}
	\caption{Schematic figure of constructing anonymous entanglement in our quantum anonymous secret sharing protocol. A sender ($Alice$) aims to share the secret to $m$ receivers ($Bobr_1, Bobr_2, \cdots, Bobr_m$) while their identities remain anonymous.}
	\label{FIG:1}
\end{figure*} 

\begin{table}[H]
\label{Protocol 3}
\begin{tabular}{@{}p{1mm}p{2mm}p{72mm}}
\toprule
\multicolumn{3}{l}{\textbf{Protocol 3:} anonymous entanglement protocol}\\
\hline
 & \multicolumn{2}{p{74mm}}{\textbf{Goal:} $|\overline{W}^{m+1}\rangle$ shared anonymously between $Alice$ and $m$ secret receivers.}\\
  & \multicolumn{2}{p{74mm}}{\textbf{Input:} $|W^{n+1}\rangle_i$ shared between $Alice$ and all potential receivers.}\\
 & \multicolumn{2}{p{74mm}}{\textbf{Output:} Anonymous entanglement success mark $E_i$.}\\
 & (1) & $Alice$ generates $|W^{n+1}\rangle_i$, separately sends $|W^{n+1}\rangle_{i,1}, |W^{n+1}\rangle_{i,2}, \cdots, |W^{n+1}\rangle_{i,n}$ to $Bob_1, Bob_2, \cdots, Bob_n$, and keeps $|W^{n+1}\rangle_{i,n+1}$.\\
 & (2) & Each non-receiver $Bob_j \notin [\mathcal{R}]$ measures in the $\mathcal{Z}$-basis and publishes his result $e_j$; receivers do not perform any measurement but publishes $e_j = 0$.\\
  & (3) & $Alice$ calculate $E_i=\Sigma_{j=1}^{n}e_j$. $E_i = 0$ means that $Alice$ and receivers successfully construct a W-state anonymous entanglement $|\overline{W}^{m+1}\rangle$.\\
\hline
\end{tabular}
\end{table}

Note that, the ($m+1$)-particle anonymous entangled W-states are of the following form:
\begin{align}
\label{eq.3}
|\overline{W}^{m+1}\rangle=&\frac{1}{\sqrt{2m}}(\sqrt{m}|\overbrace{00\cdots 01}^{m+1}\rangle + |00\cdots 10\rangle \nonumber \\
& + \cdots + |10\cdots 00\rangle ).
\end{align}    

The state $|\overline{W}^{m+1}\rangle$ is referred to as the 'perfect W state'\cite {ref26}, which is an asymmetric W state. This state is capable of facilitating flawless quantum teleportation and superdense coding. In contrast, symmetric W states are characterized by their ability to enable teleportation only with a certain probability. Fig. \ref{FIG:1} illustrates the methodology for constructing anonymous entanglement within this protocol.

\subsection{Quantum Anonymous Secret Sharing Protocol}

\begin{figure}[H]
	\centering
		\includegraphics[width=1.0\linewidth]{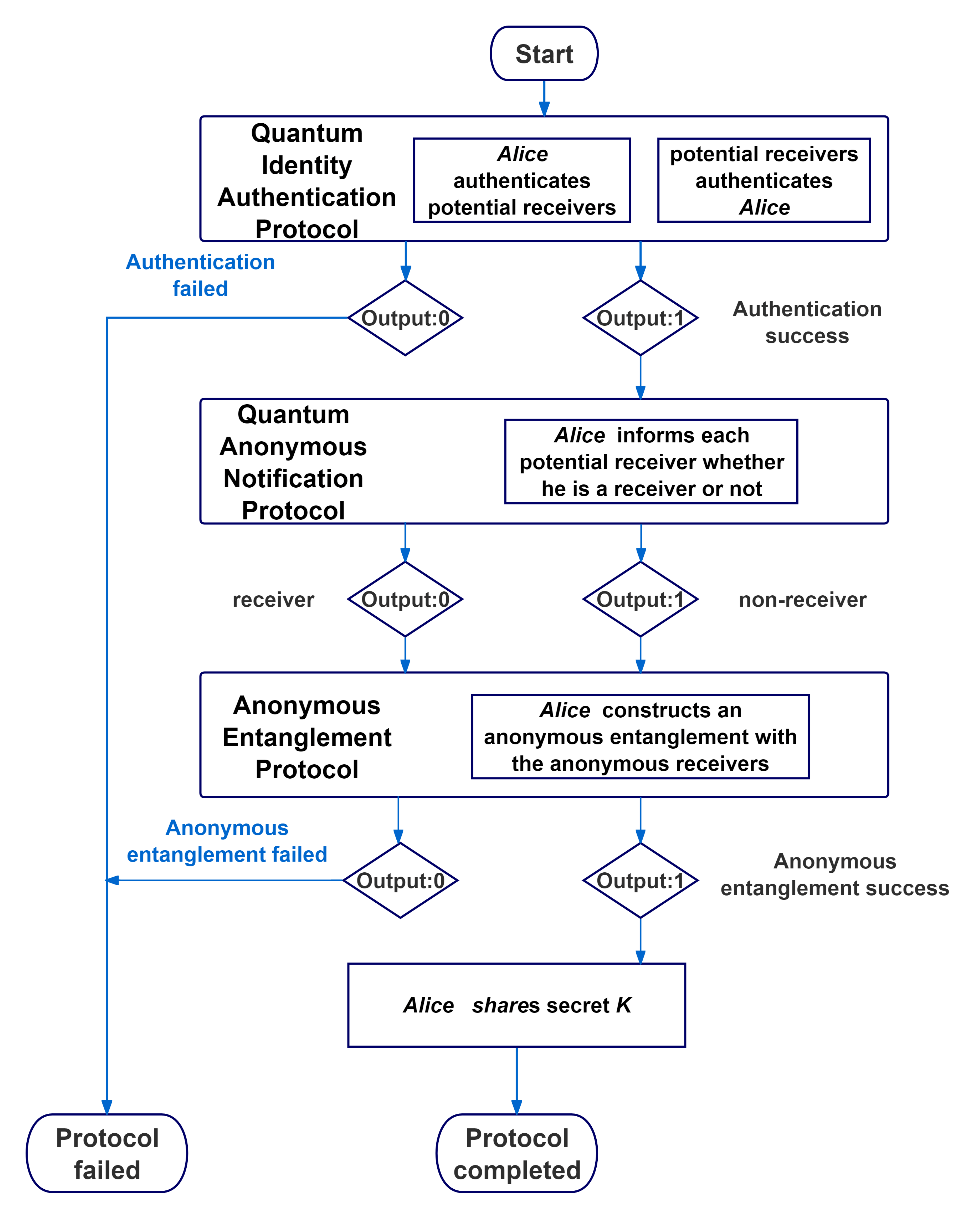}
	\caption{Flowchart of constructing an anonymous secret sharing protocol}
	\label{FIG:2}
\end{figure}

Based on the corresponding sub-protocols proposed in the previous subsections, the complete protocol for anonymously sharing $t$ bit secret $K= \{|k\rangle_1, |k\rangle_2, \cdots, |k\rangle_t$\}($|k\rangle_i = \alpha |0\rangle + \beta |1\rangle, |\alpha|^2+|\beta|^2=1, i\in[1,t]$) among $m$ anonymous receivers is given in protocol \hyperref[Protocol 4]{4}, whose flowchart is shown in Fig. \ref{FIG:2}.

Before the execution of the protocol, the corresponding preparations need to be completed. $Alice$ initiates the process by generating $x$ $|W^{n+1}\rangle$ ($x>t\cdot \frac{n-m}{2m}+n$). Out of these, $n$ states are designated as notification states, which are preprocessed as outlined in step (1) of protocol \hyperref[Protocol 2]{2}. The remaining ($x-n$) states are utilized for constructing anonymous entanglement, adhering to the distribution guidelines specified in step (2) of protocol \hyperref[Protocol 2]{2} and step (1) of protocol \hyperref[Protocol 3]{3}. $Alice$ will send all the particles at once, she puts the particles in $Q_1, Q_2, \cdots, Q_n, Q_{n+1}$ separately, $Q_j$ ($j\in [1,n+1]$) contains the $j$-th particles in each W state. After that, she prepares $n$ single photon tokens for mutual authentication ruled by protocol \hyperref[Protocol 1]{1}, step (1). Then she randomly inserts the $j$-th token corresponding to $Bob_j$ into $Q_j$, getting $Q'_j$. $Q_{n+1}$ does not need to insert the token, $Alice$ keeps this sequence in her hand. Finally, $Alice$ separately sends $Q'_1, Q'_2, \cdots, Q'_n$ to $Bob_1, Bob_2, \cdots, Bob_n$. 

When the protocol is executed, the authentication and notification of each $Bob_j$ is done in order. $Alice$ can determine a random order on her own and announce it before proceeding to the corresponding step.

\begin{table}[H]
\label{Protocol 4}
\begin{tabular}{@{}p{1mm}p{2mm}p{72mm}}
\toprule
\multicolumn{3}{l}{\textbf{Protocol 4:} anonymous secret sharing protocol}\\
\hline
 & \multicolumn{2}{p{74mm}}{\textbf{Goal:} $Alice$ shares $K=\{|k\rangle_i\}_{i=1}^t$ between $m$ anonymous receivers.}\\
 & \multicolumn{2}{p{74mm}}{\textbf{Input:} $\{PIN^{aj}, PIN^{bj}\}_{j=1}^n$; $\{Q'_j\}_{j=1}^n$ and $Q_{n+1}$ distributed in advance; secret $\{|k\rangle_i\}_{i=1}^t$.}\\
   & (1) & Identity authentication.\\
     & & \qquad $Alice$ executes identity authentication with $Bob_1, Bob_2, \cdots, Bob_n$ in order. For each $Bob_j$, she notifies the location of $l$ authentication photons in $Q'_j$, and executes protocol \hyperref[Protocol 1]{1}. If outputs $U_j=0$, the protocol\\
\hline
\end{tabular}
\end{table}

\begin{table}[H]
\begin{tabular}{@{}p{1mm}p{2mm}p{70mm}}
\toprule
\multicolumn{3}{l}{Continuation of \textbf{Protocol 4}}\\
\hline
  & & is terminated. If $U_1=U_2=\cdots=U_n=1$, the legal identity of all potential receivers is authenticated, perform the next step.\\
 & (2) & Notification. \\
& & \qquad $Alice$ broadcasts the location of $n$ notification states in $Q'_1, Q'_2, \cdots, Q'_n$, then executes protocol inform each $Bob_j$ whether he is a secret receiver or not in turn.\\
 & (3) & Anonymous entanglement.\\
& & \qquad $Alice$ and $Bob_1, Bob_2, \cdots, Bob_n$ try to establish $t$ W state anonymous entanglement for secret sharing. For the $i$-th round, $Alice$ inputs $|W^{n+1}\rangle_i$ and perform protocol \hyperref[Protocol 3]{3}. If protocol outputs $E_i=0$, then a new anonymous entanglement is established. Otherwise, execute the next round until getting $t$ anonymous entanglement.\\
 & (4) & Secret sharing.\\
& & \qquad $Alice$ use $t$ $|\overline{W}^{m+1}\rangle$ to share $t$ bits of quantum information. For the $i$-th round, $Alice$ performs a joint Bell state measurement of $|k\rangle_i$ and $|\overline{W}^{m+1}\rangle_{i,n+1}$ in her hand. She announces her measurement result $p_{i,n+1}$ (possible results contain $|\Psi^+\rangle$, $|\Psi^-\rangle$, $|\Phi^+\rangle$, $|\Phi^-\rangle$). All secret receivers measure the $|\overline{W}^{m+1}\rangle_{i,1}, |\overline{W}^{m+1}\rangle_{i,2}, \cdots, |\overline{W}^{m+1}\rangle_{i,n}$ in their hands on $\mathcal{Z}$-basis and keep the measurement results $p_{i1}, p_{i2}, \cdots, p_{im}$. This round is completed, $Alice$ turns to share the next bit.\\
\hline
\end{tabular}
\end{table}

This protocol primarily addresses the methodology for sharing quantum information. The sharing of classical information can be regarded as a specific instance within the broader context of quantum information sharing. For example, sharing $|k\rangle_i$ under the conditions where either $\alpha=0$ or $\beta=0$. Under these parameters, it becomes feasible to share classical binary information 1 or 0.

\subsection{Quantum Anonymous Secret Recovering Protocol}

\begin{figure}[h]
	\centering
		\includegraphics[width=1\linewidth]{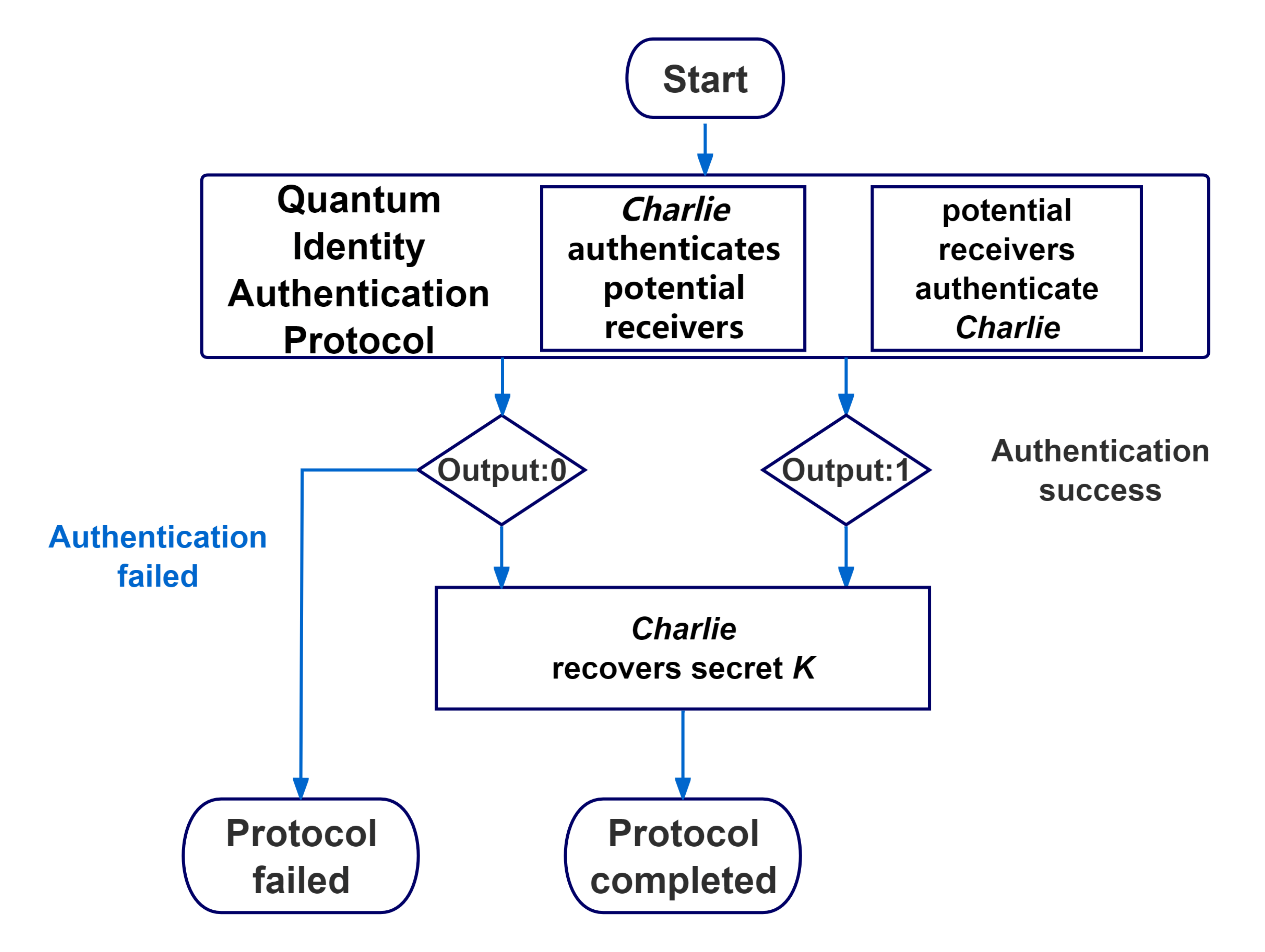}
	\caption{Flowchart of constructing an anonymous secret-recovering protocol}
	\label{FIG:3}
\end{figure}

$Charlie$ is the pre-designated secret restorer. In the secret recovering phase, he will recover the secret based on the measurement results published by $Alice$ and the information published by other potential receivers. Note that he doesn't have to know the real identity of the secret receivers in this process. Here we propose the secret recovering scheme in protocol \hyperref[Protocol 5]{5}, and the flowchart in Fig. \ref{FIG:3}. 

Before recovering, $Charlie$ generates $t$ W states $|W^{n+1}\rangle_1,$ $ |W^{n+1}\rangle_2, \cdots, |W^{n+1}\rangle_t$, following the similar preparation and distribution rule in section 3.4, but without preparing notification states. This process produces sequences $O_1, O_2, \cdots, O_n, O_{n+1}$. After that, he prepares authentication single photons token for mutual authentication with other potential receivers and randomly inserts these photons corresponding to $Bob_j$ into $O_j$, which is converted to $O'_j$. Denote his personal identity number used in this process as $\{PIN^{cj}\}_{j=1}^n$. $O_{n+1}$ does not need to insert the token, $Charlie$ keeps this sequence in his hand. Finally, $Charlie$ separately sends $O'_1, O'_2, \cdots, O'_n$ to $Bob_1, Bob_2, \cdots, Bob_n$.

Same to $Alice$, $Bob_n$ can determine a random order for authentication on his own, and announce it before proceeding to step (1).

\begin{table}[H]
\label{Protocol 5}
\begin{tabular}{@{}p{2mm}p{2mm}p{72mm}}
\toprule
\multicolumn{3}{l}{\textbf{Protocol 5:} anonymous secret recovering protocol}\\
\hline
 & \multicolumn{2}{p{74mm}}{\textbf{Goal:} $Bob_n$ recovers the secret shared by $Alice$.}\\
 & \multicolumn{2}{p{74mm}}{\textbf{Input:} $\{PIN^{bj}, PIN^{cj}\}_{j=1}^n$; $O'_1, O'_2,$$ \cdots, O'_n$ and $O_{n+1}$ distributed in advance.}\\
  & (1) & Identity authentication.\\
 & & \qquad $Charlie$ executes identity authentication with $Bob_1, Bob_2, \cdots, Bob_n$ in order. For each $Bob_j$ ($j \in [1,n]$), he notifies the location of $l$ authentication photons in $O'_j$, and executes protocol \hyperref[Protocol 1]{1}. If the protocol outputs $U_j=0$, terminate the protocol and consider $Bob_j$ illegal. If $U_1=U_2=\cdots=U_n=1$, the legal identity of all potential receivers is authenticated, perform the next step.\\
  & (2) & Secret recovering.\\
 & & \qquad $Charlie$ use $t$ $|W^{n+1}\rangle$ to recover $t$ bits of quantum information. For the $i$-th round, secret receivers perform a unitary operation based on the measurement result in protocol \hyperref[Protocol 4]{4}, step (4). Take $Bobr_j$ as an example, if $p_{ij}=0$, he performs an $I$ operation; if $p_{ij}=1$, he performs an $X$ operation. Then, $Bob_1, Bob_2, \cdots, Bob_n$ measure the particles in their hands on $\mathcal{Z}$-basis and report their results $p'_{i1}, p'_{i2}, \cdots, p'_{in}$ to $Charlie$ in order. $Charlie$ can perform corresponding unitary operations on his particle according to $p_{i,n+1}$ and $p'_1, p'_2, \cdots, p'_n$ to get $|k\rangle_i$, the operation rules are shown in Table \ref{Table 3}.\\
  \hline
\end{tabular}
\end{table}

\begin{table*}[t]
\caption{Rules for $Charlie$ to perform unitary operations.}\label{Table 3}
\centering
\begin{tabular}{|c|c|c|c|c|}
\toprule
$p_{i,n+1}$ & \multicolumn{2}{c|}{$|\Psi^+\rangle$} & \multicolumn{2}{c|}{$|\Psi^-\rangle$}\\
\hline
$p'_{i1}, p'_{i2}, \cdots, p'_{in}$ & $MR_1$ & $MR_2$ & $MR_1$ & $MR_2$\\
$|W^{n+1}\rangle_{i,n+1}$ & $\alpha |1\rangle + \beta |0\rangle$ & $\alpha |0\rangle + \beta |1\rangle$ & $\alpha |1\rangle - \beta |0\rangle$ & $\alpha |0\rangle - \beta |1\rangle$\\
$Charlie$'s unitary operation & $X$ & $I$ & $Y$ & $Z$\\
\hline
$p_{i,n+1}$ & \multicolumn{2}{c|}{$|\Phi^+\rangle$} & \multicolumn{2}{c|}{$|\Phi^-\rangle$}\\
\hline
$p'_{i1}, p'_{i2}, \cdots, p'_{in}$ & $MR_1$ & $MR_2$ & $MR_1$ & $MR_2$\\
$|W^{n+1}\rangle_{i,n+1}$ & $\alpha |0\rangle + \beta |1\rangle$ & $\alpha |1\rangle + \beta |0\rangle$ & $\alpha |0\rangle - \beta |1\rangle$ & $\alpha |1\rangle - \beta |0\rangle$\\
$Charlie$'s unitary operation & $I$ & $X$ & $Z$ & $Y$\\
\hline
\end{tabular}
\end{table*}

If $Bob_1, Bob_2, \cdots, Bob_n$ execute the protocol honestly, there are two possible measurement results for $p'_{i1}, p'_{i2}, \cdots, p'_{in}$. One is that ($n-1$) potential receivers' measurement results are 0 and one potential receiver's measurement result is 1, we denote it as "measurement result 1", or "MR1" for short. The other is that $n$ potential receivers' measurement results are all 0, we denote it as "measurement result 2", or "MR2" for short.

\section{Correctness and security}

As delineated in Definition 1, the primary aim of QASS is to ensure recover-ability, recover-security, and recover-anonymity. The core of QASS is anonymous entanglement. It is pertinent to note that the construction of anonymous W state entanglement within this framework is probabilistic. Specifically, there is a possibility of construction failure of the W state when $E_i \neq 0$ in protocol \hyperref[Protocol 3]{3}. The first subsection will focus on elucidating the probability of successful anonymous entanglement, quantified as a function of parameters $m$ and $n$ in the network. This will be followed by an analysis of the protocol's correctness and security.

\subsection{Entanglement Probability}

\textbf{Theorem 1} (Probability of successful anonymous entanglement) \textbf{.} In a noise-free channel, assume a sender, denoted as $Alice$, aims to establish anonymous entanglement with $m$ undisclosed receivers utilizing the state $|W^{n+1}\rangle$. Within this system, there are $n$ potential receivers, and all parties involved adhere to honest protocols. Under these conditions, the probability of successfully generating an anonymous entangled state, represented as $|\overline{W}^{m+1}\rangle_i$, is calculated to be $\frac{2m}{n+m}$.

\textit{Proof.} Let $|\Vec{0}\rangle \langle \Vec{0}|_{n-m}$ denote the projection on the $|0\rangle$ state of $(n-m)$ non-receivers. The probability $P_{|\overline{W}^{m+1}\rangle_i}$ of obtaining this state can be expressed as 
\vspace{-1em}

\begin{align}
    \label{eqn.5}
P_{|\overline{W}^{m+1}\rangle_i} =& Tr[|W\rangle \langle W|_{n+1} \cdot (I_A\otimes I_{m+1}\otimes |\Vec{0}\rangle \langle \Vec{0}|_{n-m})] \nonumber \\
=& \frac{2m}{n+m}.
\end{align}

Theorem 1 states that in the honest implementation, the probability of successful anonymous entanglement in protocol 3 is based on the proportion of the number of receivers and non-receivers. The success rate is higher when the number of secret receivers is large.

\subsection{Protocol Correctness}

\textbf{Theorem 2} (Correctness of secret sharing) \textbf{.} In a noise-free channel, provided that all participants act honestly and protocol \hyperref[Protocol 4]{4} proceeds without termination, the objective of distributing the secret quantum state $|k\rangle_i$ in an anonymous manner is achieved with precision.

\textit{Proof.} We examine the accuracy of the secret sharing protocol, including an assessment of the individual sub-protocols. During the initial phase, designated as step (1), $Alice$ runs protocol \hyperref[Protocol 1]{1} with each potential receiver separately to finish authentication. Specifically, in an authentication sequence involving $Bob_i$, $Alice$ prepares a quantum state denoted as $q_i$, applies a unitary operation $U_i$, and transmits the transformed state $q_i'=U_i \cdot q_i$ to $Bob_i$. Based on the properties of unitary transformation, for any unitary transformation $U$, it follows $U\cdot U^T=1$. In other words, for any unitary transformation $U$, if $U|\phi \rangle=|\psi \rangle$, then it must hold that $U|\psi \rangle=|\phi \rangle$. Consequently, if $Bob_i$ possesses a legal identity, he will apply $U_i$ to receive $q_i''=U_i \cdot q_i'=q_i$. This allows both parties to verify the authenticity of each other's identity.

In step (2), the potential receivers are notified one after another anonymously according to protocol \hyperref[Protocol 2]{2}. The notification state shared by them is obtained by applying several $X$ operators on $|W^{n+1}\rangle$. If $Alice$ selects $Bob_i$ as one of her unique receivers, the number of $X$ operators is random but odd. $Bob_i$ does not publish his measurements, so only he knows his receiver identity by calculating $N_i$ after others publish their measurement results. 

The analysis of step (3) and protocol \hyperref[Protocol 3]{3} follows the correctness of the anonymous entanglement protocol presented in Ref. \cite{ref21}, which provides a method for constructing anonymous entangled EPR pairs. Differently, we aim to build anonymous W state entanglement instead of EPR pairs between multiple participants. If we partition the $|W^{n+1}\rangle$ state depicted in Eq.(\ref{eqn.1}) into two subsystems in the way: $i/(1,\cdots,i-1,i+1,\cdots,n+1)$, then when $i\neq n+1$ it can be rewritten in the Schmidt decomposition form:
\begin{align}
\label{Eqnarray 3}
|W^{n+1}\rangle = &\sqrt{\frac{n+m-1}{n+m}} |\psi ^n\rangle |0\rangle _i \nonumber \\
&+ \frac{1}{\sqrt{n+m}} |00\cdots 0\rangle |1\rangle _i,
\end{align}
where
\begin{align}
|\psi ^n\rangle =& \frac{1}{\sqrt{n+m-1}} (\sqrt{m}|\overbrace{00\cdots 01}^{n}\rangle \nonumber \\
&+ |00\cdots 10\rangle + \cdots + |10\cdots 00\rangle ).
\end{align}  

Thus, after $Bob_i$'s $\mathcal{Z}$-basis measurement on particle $i$, the total state will collapse into $|\psi ^n\rangle$. If we decompose the state in the same way in the case that the measurement results of all non-receivers are all 0, then the total state will collapse into $|\overline{W}^{m+1}\rangle_i$. Then $Alice$ can perfectly transmit a quantum state to the receivers. 

In step (4), $Alice$ conducts a Bell State measurement on the quantum states $|k\rangle_i$ and $|\overline{W}^{m+1}\rangle_{i,n+1}$. Subsequently, she discloses the measurement outcome. Following this, $Bobr_1, Bobr_2, \ldots, Bobr_m$ proceed to measure their own particles using the $\mathcal{Z}$-basis. Upon the completion of these measurements, the process of quantum teleportation is considered finalized. The entangled state resulting from these operations can be reformulated as:
\begin{align}
&|k\rangle_i |\overline{W}^{m+1}\rangle_i \\
=& (\alpha |0\rangle + \beta |1\rangle)\frac{1}{\sqrt{2}} (|\Tilde{W}^m\rangle |0\rangle +|00\cdots 0\rangle |1\rangle)\nonumber \\
=& \frac{1}{\sqrt{2}}(\alpha |00\rangle |\Tilde{W}^m\rangle + \alpha |01\rangle |00\cdots 0\rangle \nonumber \\
&+ \beta |10\rangle |\Tilde{W}^m\rangle + \beta |11\rangle |00\cdots 0\rangle) \nonumber \\
=& \frac{1}{2} [|\psi ^+\rangle (\alpha |\Tilde{W}^m\rangle + \beta |00\cdots 0\rangle) \nonumber \\ 
&+ |\psi ^-\rangle (\alpha |\Tilde{W}^m\rangle - \beta |00\cdots 0\rangle)\nonumber \\
&+ |\phi ^+\rangle (\alpha |00\cdots 0\rangle + \beta |\Tilde{W}^m\rangle) \nonumber \\
&+ |\phi ^-\rangle (\alpha |00\cdots 0\rangle - \beta |\Tilde{W}^m\rangle)],\nonumber
\end{align} 
where
\begin{small}
        \begin{align}
|\Tilde{W}^m\rangle = \frac{1}{\sqrt{m}}(|0\cdots 01\rangle + |0\cdots 10\rangle + \cdots + |10\cdots 0\rangle ).
\end{align}
\end{small}
which is a $m$-particle symmetric W state.

After $\mathcal{Z}$-basis measurement, each receiver acquires an equitable portion of the confidential information. This action culminates in the substantiation of Theorem 2. Subsequently, the validation of protocol \hyperref[Protocol 5]{5} is established through the ensuing result.

\textbf{Theorem 3} (Correctness of secret recovering) \textbf{.} If all participants adhere to the protocol with integrity and protocol \hyperref[Protocol 5]{5} proceeds without termination, the secret restorer is capable of achieving anonymous recovery of the quantum information denoted by $|k\rangle_i$.

\textit{Proof.} Step (1) adheres to the guidelines established in protocol \hyperref[Protocol 1]{1}; therefore, the verification of its accuracy is aligned with the relevant section in the proof of Theorem 2. During step (2), $Bob_n$ employs one of the unitary operators ($I$, $X$, $Y$, $Z$) to transform his particle into the state ‘$|k\rangle_i$’. This process is described below, which is divided into four cases, each contingent upon the variance in $Alice$'s measurement result:

(1) $Alice$'s BSM result is $|\psi ^+\rangle$.
\begin{small}
    \begin{align}
\label{eq.1}
&|W^{n+1}\rangle = \frac{1}{\sqrt{2}} (|\Tilde{W}^n\rangle |0\rangle +|00\cdots 0\rangle |1\rangle) \\
&= \frac{1}{\sqrt{2}} [\frac{1}{\sqrt{2}} (|\Tilde{W}^{n-1}\rangle |0\rangle +|00\cdots 0\rangle |1\rangle) |0\rangle+|00\cdots 0\rangle |1\rangle]\nonumber \\
&|W^{n+1}\rangle \to (\alpha I^{\otimes m-1}\otimes X + \beta I^{\otimes m})|W^{n+1}\rangle \\
&|W^{n+1}\rangle = \frac{1}{2} [|\Tilde{W}^{n-1}\rangle (\alpha |1\rangle + \beta |0\rangle) + |00\cdots 0\rangle \nonumber \\
& (\alpha |0\rangle  + \beta |1\rangle)]|0\rangle +\frac{1}{\sqrt{2}}|00\cdots 0\rangle (\alpha |1\rangle + \beta |0\rangle) |1\rangle \\
&= \frac{1}{\sqrt{2}}[|\Tilde{W}^n\rangle (\alpha |1\rangle + \beta |0\rangle)+ |00\cdots 0\rangle (\alpha |0\rangle + \beta |1\rangle)]\nonumber
\end{align}
\end{small}

(2) $Alice$'s BSM result is $|\psi ^-\rangle$.
\begin{small}
    \begin{align}
         |W^{n+1}\rangle \to & (\alpha I^{\otimes m-1}\otimes X - \beta I^{\otimes m})|W^{n+1}\rangle \\
         |W^{n+1}\rangle =& \frac{1}{\sqrt{2}}[|\Tilde{W}^n\rangle (\alpha |1\rangle - \beta |0\rangle)+ |00\cdots 0\rangle (\alpha |0\rangle - \beta |1\rangle)] 
    \end{align}
\end{small}

(3) $Alice$'s BSM result is $|\phi ^+\rangle$.
\begin{small}
    \begin{flalign}
        |W^{n+1}\rangle \to & (\alpha I^{\otimes m} + \beta I^{\otimes m-1}\otimes X)|W^{n+1}\rangle \\
        |W^{n+1}\rangle=& \frac{1}{\sqrt{2}}[|\Tilde{W}^n\rangle (\alpha |0\rangle + \beta |1\rangle) + |00\cdots 0\rangle (\alpha |1\rangle + \beta |0\rangle)]
    \end{flalign}
\end{small}

(4) $Alice$'s BSM result is $|\phi ^-\rangle$.
\begin{small}
    \begin{flalign}
    \label{eq.2}
        |W^{n+1}\rangle \to & (\alpha I^{\otimes m} - \beta I^{\otimes m-1}\otimes X)|W^{n+1}\rangle \\
        |W^{n+1}\rangle =& \frac{1}{\sqrt{2}}[|\Tilde{W}^n\rangle (\alpha |0\rangle - \beta |1\rangle) + |00\cdots 0\rangle (\alpha |1\rangle - \beta |0\rangle)]
    \end{flalign}
\end{small}

Here, $|\Tilde{W}^n\rangle$ corresponds to the case of $MR1$ in table \ref{Table 3}; $|00\cdots 0\rangle$ corresponds to the case of $MR2$ in table \ref{Table 3}. It follows from Eq. (\ref{eq.1})-(\ref{eq.2}) that the particle in $Charlie$'s hand is the same as revealed in table \ref{Table 3}. Thus $Charlie$ can obtain $|k\rangle_i$ after performing the corresponding operation according to the rules of table \ref{Table 3}. This completes the proof of Theorem 3.

\subsection{Authentication Security}

Within the scope of our security framework, we consider the presence of the active adversary. This entity is capable of executing any quantum operation and may target some participants in the system, as discussed in the referenced literature\cite{ref27}.

Regarding the identity authentication protocol, the utilization of a one-time $P\hspace{-0.5mm}I\hspace{-0.5mm}N$-based token renders any attempt by an adversary to intercept this token futile. The robustness of the single-photon Quantum Identity Authentication (QIA) protocol has been rigorously analyzed and its resilience against various attack methodologies has been affirmed\cite{ref28}. Consequently, our analysis primarily focuses on the scenario where an active adversary attempts to impersonate a designated receiver, $Bob_j$, or the sender, $Alice$. We denote $\mathcal{D}$ as the subset comprising these active adversaries. In cases of impersonating $Alice$, let $\mathcal{W}^1$ represent the adversaries’ source of quantum single photons, generated independently of the legitimate $P\hspace{-0.5mm}I\hspace{-0.5mm}N^{aj}$ and $P\hspace{-0.5mm}I\hspace{-0.5mm}N^{bj}$. The probability of an adversary successfully passing the authentication process is then quantified as
\begin{align}
P_{pass1}[\mathcal{C}, \mathcal{W}^1] = \frac{1}{2^l}.
\end{align}

When impersonating $Bob_i$, let $\mathcal{W}^2$ denote the adversaries’ quantum register of the state distributed by $Alice$; $\mathcal{U}^2$ denote the random unitary operation since adversaries operate without the true $P\hspace{-0.5mm}I\hspace{-0.5mm}N^b$. Then, the probability of an adversary passing authentication is given by
\begin{align}
    P_{pass2}[\mathcal{C}, \mathcal{W}^2, \mathcal{U}^2] = \frac{1}{2^l}.
\end{align}

Thus, for $l$ large enough, it can be considered that $P_{pass1}=P_{pass2}\approx 0$. Therefore, it can be considered that the adversary cannot pass the identity authentication, and the authentication part can ensure its security.

\subsection{Receiver Anonymity}

It should be noted that the honesty of potential receivers does not preclude the possibility of the malicious adversary obstructing the shared secret between $Alice$ and secret receivers. Consequently, the reliability of both protocol \hyperref[Protocol 4]{4} and protocol \hyperref[Protocol 5]{5} is vulnerable to such malicious interventions. This issue could be addressed through the implementation of quantum message authentication techniques. A pertinent question arises: does this approach compromise the anonymity of the secret receivers? In the subsequent analysis, it is demonstrated that our protocols maintain receiver anonymity. Even in scenarios where the adversary controls some dishonest potential receivers, the anonymity of the receivers remains intact.

\textbf{Theorem 4} (Receiver anonymity in the active adversary scenario) \textbf{.} Consider the noise-free perfect channel, our quantum anonymous secure sharing protocol with W states, is receiver-anonymous in the active adversary scenario.

\textit{Proof.} In section 2, we introduce the security definition of the guessing probability in Eq.(\ref{eqn.2}). In our protocol, for the ideal case, $P[br_i = b_j|br_i\notin \mathcal{D}]$ should be $\frac{m_\mathcal{H}}{n-|\mathcal{D}|}$. Here $m_\mathcal{H}$ represents the number of honest receivers. Besides, the \textit{guessing probability}\cite{ref21}, in our QASS protocol is
\begin{align}
    &P_{guess}[br_i|\mathcal{W}^{\mathcal{D}}, \mathcal{C}, br_i\notin \mathcal{D}]\\ 
    = & \mathop{max}\limits_{M^j} \sum\limits_{b_j\in \mathcal{H}} P[br_i = b_j|br_i\notin \mathcal{D}]Tr[M^j \cdot \rho_{\mathcal{W}^{\mathcal{D}}, \mathcal{C}|br_i = b_j}],\nonumber
\end{align}
where the guessing probability is the maximum taken over the set of positive operator-valued measures {$M^j$} for the adversaries, and $\rho_{\mathcal{W}^{\mathcal{D}}, \mathcal{C}|br_i = b_j}$ is the reduced quantum state of dishonest participants at the end of the protocol given that $Bob_j$ is the receiver $Bobr_i$. The premise of achieving receiver security is that the adversary cannot distinguish the honest non-receiver from the receiver.

We prove anonymity for all involved sub-protocols separately. The outcome of Protocol \hyperref[Protocol 2]{2} confidentially informs each potential receiver of their status as a receiver or not, without divulging additional information. The adversary's reduced quantum state upon completion of this protocol remains uncorrelated with the identity of the receiver. Specifically, for any $Bob_i$ not included in the subset $\mathcal{D}$ of adversaries, the condition $\rho _{\mathcal{W}^{\mathcal{D}}, \mathcal{C}, br_i}=\rho {\mathcal{W}^{\mathcal{D}}, \mathcal{C}}$ holds true. In practical terms, this implies that in scenarios where the adversary controls all entities except for $Bobr_i$ and acquires the measurement outcomes of the notification state, the probability of correctly guessing the identity of $Bobr_i$, denoted as $P{guess}[Bobr_i]$, remains at 1/2. This satisfies Eq.(\ref{eqn.2}), that the receiver identity about $Bobr_i$ remains inaccessible to the adversary.

An active adversary might target all quantum sequences to ascertain whether $Bobr_i$ is the intended receiver. While Protocol \hyperref[Protocol 2]{2} does not solely thwart such an attack, in Protocol \hyperref[Protocol 4]{4}, $Alice$ reveals the positions of all notification states only after successful authentications. If the adversary indiscriminately compromises all particles, this mode of attack will be detected during the authentication phase, a scenario substantiated in our analysis of Protocol \hyperref[Protocol 1]{1}. Furthermore, selectively attacking a specific notification particle is not feasible. To elucidate this, we introduce $P_{attack}$:
\begin{align}
    P_{attack}=\frac{n}{x+l}.
\end{align}

Thus, for $x$ and $l$ large enough, it can be considered that $P_{attack}\approx 0$. In other words, it is impossible for the adversary to only attack the designated notification state particles to obtain the identity information of a receiver by guessing.

If some or all potential receivers except $Bobr_i$ are governed by an active adversary, the worst case would be that the parity of broadcast results changes from even to odd or vice versa, which prevents the receiver from being notified or makes the sender aware of
the presence of an adversary. Nevertheless, it reveals no information on the identities of $Bobr_i$. In summary, we have progressively analyzed the possibilities of various types of adversaries and proved the proposed quantum protocol is perfectly receiver-secure.

In steps (3) or (4), each participant performs local operations and measurements in sequence. The key to the protocol to ensure anonymity is to establish secure anonymous entanglement. To achieve this, a feasible prerequisite is to guarantee that the measurement results published by non-receivers are random and indistinguishable. Therefore, in our protocol, this is equivalent to
\begin{align}
\label{eqn.6}
& P_{measure}[b_i|\mathcal{W}^{\mathcal{D}}, \mathcal{C}, b_i\notin \mathcal{D}] \nonumber \\
=& P_{measure}[b_j|\mathcal{W}^{\mathcal{D}}, \mathcal{C}, b_j\notin \mathcal{D}],
\end{align}
where $P_{measure}$ represents the probability of a possible measurement result, $i \neq j$. According to the conditions for the successful anonymous entanglement, we consider its probability of 0.

Considering the presence of active adversaries, the shared W state in the protocol should be
\begin{small}
    \begin{align}
    &|\omega_{n+1}\rangle =\frac{1}{\sqrt{n+m}}(\sqrt{m}|00\cdots 01\rangle_{A\mathcal{H}} \otimes|\varphi_0\rangle_{\mathcal{D}}+ \\ 
    &\quad |00\cdots 10\rangle_{A\mathcal{H}} \otimes|\varphi_1\rangle_{\mathcal{D}}+ \cdots +
    |10\cdots 00\rangle_{A\mathcal{H}} \otimes|\varphi_n\rangle_{\mathcal{D}} ).\nonumber 
    \end{align}
\end{small}

Denote $\sqrt{m}|00\cdots 01\rangle_{A\mathcal{H}}$ as $|\psi_0\rangle$, $|00\cdots 10\rangle_{A\mathcal{H}}$ as $|\psi_1\rangle$, $\cdots$, $\sqrt{m}|10\cdots 00\rangle_{A\mathcal{H}}$ as $|\psi_n\rangle$, then 
\begin{align}
        |\omega_{n+1}\rangle=\frac{1}{\sqrt{n+m}}\sum_{x=0}^n (|\psi_x\rangle \otimes |\varphi_x\rangle).
\end{align}

By tracing out $P_{measure}[b_i]$, after measurement,
\begin{align}
\label{eqn.7}
& P_{measure}[b_i]= Tr_{n-m}[(|\omega_{n+1}\rangle \langle \omega_{n+1}|)\cdot (I_{n} \otimes |\Vec{0}\rangle \langle \Vec{0}|_{i})] \nonumber \\
& \quad = \frac{1}{n+m} \sum_{y=0}^n \sum_{z=0}^n Tr [|\psi_y\rangle \langle \psi_z|]Tr[|\varphi_y\rangle \langle \varphi_z|] \nonumber \\
& \qquad \cdot (I_{n} \otimes |\Vec{0}\rangle \langle \Vec{0}|_{i})].
\end{align}

Since the $I$ operation does not change the trace, we can rewrite and simplify Eq. (\ref{eqn.7}) as 
\begin{align}
P_{measure}[b_i]=& \frac{1}{n+m}[(m \langle \Vec{0}_{i}|\varphi_0\rangle)+ \sum_{x=1}^{j-1} (\langle \Vec{0}_{i}|\varphi_x\rangle) \nonumber \\
& + \sum_{x=j+1}^{n} (\langle \Vec{0}_{i}|\varphi_x\rangle)] \nonumber \\
=& \frac{n+m-1}{n+m}.
\end{align}

Therefore, $P_{measure}[b_i]$ is independent of the identity of $Bob_i$, or $P_{measure}[b_i]=P_{measure}[b_j]$, Eq.(\ref{eqn.6}) holds. Then we can calculate 
\begin{flalign}
        &P_{guess}[br_i|\mathcal{W}^{\mathcal{D}}, \mathcal{C}, br_i\notin \mathcal{D}]\\ 
        = & \mathop{max}\limits_{M^j} \sum\limits_{b_j\in \mathcal{H}} P[br_i = b_j|br_i\notin \mathcal{D}]Tr[M^j \cdot \rho_{\mathcal{W}^{\mathcal{D}}, \mathcal{C}|br_i = b_j}]\nonumber \\
        \leq & \mathop{max}_j P[br_i = b_j|br_i\notin \mathcal{D}]Tr[\sum\limits_{b_j\in \mathcal{H}} M^j \cdot \rho_{\mathcal{W}^{\mathcal{D}}, \mathcal{C}}] \nonumber \\
        = & \mathop{max}_j P[br_i = b_j|br_i\notin \mathcal{D}].
\end{flalign}

It can be seen that the guessing probability $P_{guess}$ satisfies our anonymity requirement in Definition 2. However, due to the attacks from malicious potential receivers, their broadcast results would be changed, which causes protocol \hyperref[Protocol 4]{4} to abort or pass. Even so, no adversary obtains any information about the identity of the receivers, since all honest potential receivers exhibit the same. Thus, the anonymity of the receivers is guaranteed regardless of how many potential receivers are controlled by the active adversary. But we remark that the malicious parties can prevent $Alice$ and receivers from sending and sharing the desired secret. For example, the dishonest parties can measure the W state on a different basis affecting the resulting anonymous entanglement. In this sense, protocol \hyperref[Protocol 4]{4} is not robust to malicious attacks. The reliability of this part can be ensured by quantum message authentication. Thus, even in the presence of dishonest parties, the anonymity of receivers is preserved. So Theorem 4 is proven.

The anonymity of protocol \hyperref[Protocol 5]{5} is similar to that of protocol \hyperref[Protocol 4]{4}. The known recovering W state is distributed to all participants, and the secret receiver behaves the same as the non-secret receiver except for the local measurement and unitary operation, and only transmits the measurement results through a secure classical channel with the secret restorer. Thus the completion of protocol \hyperref[Protocol 5]{5} does not break the recipient's anonymity either.

\subsection{Secret Security}

\textbf{Theorem 5} (Secret security in the active adversary scenario) \textbf{.} Consider the noise-free perfect channel, our quantum anonymous secure sharing protocol with W states, can protect secret security in the active adversary scenario.

\textit{Proof.} In protocol \hyperref[Protocol 4]{4}, the security of the secret sharing part can be guaranteed by the authentication mentioned above. The authentication of the potential receivers' identity ensures that the illegal external adversary cannot obtain the secret information. Therefore, in this subsection, we will focus on the case where the adversary controls the dishonest potential receivers. 

Considering that $Alice$ uses quantum teleportation based on perfect W state to share the quantum information $|k\rangle_i$, which is information-theoretic secure\cite{ref26}. So it is not practical to launch the attack in this step. A feasible way is in step (3), to try to join the anonymous entanglement $|\overline{W}^{m+1}\rangle_i$. There are two possibilities. The first is to make a measurement and publish the wrong result. That is, publish 0 when the measurement is 1. Only one dishonest potential receiver can publish a false measurement result to complete step (3). Otherwise, there will be a measurement result that violates the property of W state and will be found.  Suppose the dishonest potential receiver is $Bob_j$. Follow Eq.(\ref{Eqnarray 3}), in this case, the actual shared anonymous entangled state is given by
\begin{align}
\label{eq.4}
|\phi^{m+2}\rangle_i = |00\cdots 0\rangle |1\rangle.
\end{align}

Eq.(\ref{eq.4}) shows that the final shared $|\phi^{m+2}\rangle_i$ is a direct product state instead of an entangled state. Consequently, teleportation is not achievable, leading to the failure of anonymous secret sharing. Thus, the adversary naturally cannot recover the shared secret information of $Alice$.

The second possibility is not to measure but to complete steps (3) and (4) in the same way as a secret receiver. In this case, the actual shared anonymous entangled state is given by
\begin{align}
    |W^{m+1+d}\rangle_i=&\frac{1}{\sqrt{2m+d}}(\sqrt{m}|00\cdots 01\rangle + |00\cdots 10\rangle \nonumber \\
    &+ \cdots + |10\cdots 00\rangle ),
\end{align}
where $d$ denotes the number of dishonest non-receivers. This state is not a perfect W state, and the teleportation is a probabilistic success, which is detailed in \cite{ref29}. The combined state can be rewritten as
\begin{align}
&|k\rangle_i |W^{m+1+d}\rangle_i \\
=& (\alpha |0\rangle + \beta |1\rangle)\frac{1}{\sqrt{2m+d}} \nonumber \\
&(\sqrt{m}|\Tilde{W}^{m+d}\rangle |0\rangle +\sqrt{m+d}|00\cdots 0\rangle |1\rangle)\nonumber \\
=& \frac{1}{\sqrt{2}}(\alpha \sqrt{m}|00\rangle |\Tilde{W}^{m+d}\rangle + \alpha \sqrt{m+d}|01\rangle |00\cdots 0\rangle \nonumber \\
&+ \beta \sqrt{m}|10\rangle |\Tilde{W}^{m+d}\rangle + \beta \sqrt{m+d}|11\rangle |00\cdots 0\rangle) \nonumber \\
=& \frac{1}{2} [|\psi ^+\rangle (\alpha \sqrt{m}|\Tilde{W}^{m+d}\rangle + \beta \sqrt{m+d}|00\cdots 0\rangle)  \nonumber \\
&+ |\psi ^-\rangle (\alpha \sqrt{m}|\Tilde{W}^{m+d}\rangle - \beta \sqrt{m+d}|00\cdots 0\rangle)\nonumber \\
&+ |\phi ^+\rangle (\alpha \sqrt{m+d}|00\cdots 0\rangle + \beta \sqrt{m}|\Tilde{W}^{m+d}\rangle)  \nonumber \\
&+ |\phi ^-\rangle (\alpha \sqrt{m+d}|00\cdots 0\rangle - \beta \sqrt{m}|\Tilde{W}^{m+d}\rangle)].\nonumber
\end{align} 

This results in the secret restorer getting a wrong state in protocol \hyperref[Protocol 5]{5}, that is, failing to recover $|k\rangle_i$. We take one of these cases as an example, where $Alice$'s BSM result is $|\psi ^+\rangle$. Assuming that subsequent steps execute normally, the secret restorer has already distributed $|W^{n+1}\rangle$ and attempted to recover $|k\rangle_i$. 

\begin{align}
&|W^{n+1}\rangle = \frac{1}{\sqrt{2}} (|\Tilde{W}^n\rangle |0\rangle +|00\cdots 0\rangle |1\rangle) \\
&\quad = \frac{1}{\sqrt{2}} [\frac{1}{\sqrt{2}} (|\Tilde{W}^{n-1}\rangle |0\rangle +|00\cdots 0\rangle |1\rangle) |0\rangle \nonumber \\
&\qquad +|00\cdots 0\rangle |1\rangle]\nonumber \\
&|W^{n+1}\rangle \to (\alpha \sqrt{m} I^{\otimes m-1}\otimes X \nonumber \\
&\qquad + \beta \sqrt{m+d} I^{\otimes m})|W^{n+1}\rangle \\
&|W^{n+1}\rangle = \frac{1}{2} [|\Tilde{W}^{n-1}\rangle (\alpha \sqrt{m} |1\rangle + \beta \sqrt{m+d} |0\rangle) \nonumber \\
& + |00\cdots 0\rangle (\alpha \sqrt{m} |0\rangle  + \beta \sqrt{m+d} |1\rangle)]|0\rangle \nonumber \\
& +\frac{1}{\sqrt{2}}|00\cdots 0\rangle (\alpha \sqrt{m} |1\rangle + \beta \sqrt{m+d} |0\rangle) |1\rangle \\
&= \frac{1}{\sqrt{2}}[|\Tilde{W}^n\rangle (\alpha \sqrt{m} |1\rangle + \beta \sqrt{m+d} |0\rangle) \nonumber \\
& + |00\cdots 0\rangle (\alpha \sqrt{m} |0\rangle + \beta \sqrt{m+d} |1\rangle)]\nonumber
\end{align}

Therefore, when the measurements of other potential receivers are MR1, the particles held in $Charlie$'s hand is $\alpha \sqrt{m} |1\rangle + \beta \sqrt{m+d} |0\rangle$, instead of $\alpha |1\rangle + \beta |0\rangle$. $Charlie$ does not know this, so after he follows the rule and performs the unitary operation, the resulting particle is $\alpha \sqrt{m} |0\rangle + \beta \sqrt{m+d} |1\rangle$. The adversary cannot obtain more information than $Charlie$, and even $Charlie$ himself cannot get the correct secret, so the adversary cannot obtain the secret illegally by using this attack method.

In addition, it is also possible that a dishonest secret receiver will try to obtain the secret shares of other receivers. But this is not realistic in our protocol, because the measurement of the anonymous entangled particles is done locally, and there is no possibility of being attacked within the considered category.

In protocol \hyperref[Protocol 5]{5}, we assume that at least the secret retriever is honest since he is already able to have all participants' $P\hspace{-0.5mm}I\hspace{-0.5mm}N$ and the full secret. In a real scenario, this participant may be a public trusted control center that assists anonymous receivers in recovering the final secret according to their additional information. There are two kinds of channels used in protocol \hyperref[Protocol 5]{5}. One is the secure classical channel, which is used to transmit the measurement results. The second is the quantum channel, which is used to distribute the recovery W states. The security of the quantum channel is also guaranteed by the randomly inserted identity authentication single photon, and the analysis of this part is similar to that in protocol \hyperref[Protocol 4]{4}. So the attack on the channel cannot obtain valid information about the secret. In summary, we analyze the possible attack means of the adversary and exclude the possibility of a successful attack, so Theorem 5 is proved.

\subsection{Secret Integrity}

In secret sharing, the integrity of information is also a part of its security. It should be mentioned that in our protocol, the secret restorer can correctly recover $Alice$'s shared secret, but the authenticity of the secret cannot be guaranteed. Whether to give him the ability to authenticate messages depends on the requirements of real applications, and the way to do this is in the form of quantum message authentication, which we will describe shortly.

To implement message authentication, $Alice$ creates several instances of Bell state $|\Phi ^+\rangle$. She keeps one qubit of each pair and calls $\gamma$ as the other qubit. In protocol \hyperref[Protocol 4]{4}, before step (4) (secret sharing), $Alice$ creates a random classical key $\delta$, and computes $\gamma '=\textbf{authenticate}(\gamma,\delta)$. After that, she performs a teleportation measurement on $\gamma '$ using the anonymous entanglement W states generated in step (3). 

After secret recovering in protocol \hyperref[Protocol 5]{5}, $Alice$ can use an anonymous communication protocol to send $\delta$ and the teleportation bits to the restorer $Bob_n$. $Bob_n$ completes the teleportation and computes $\gamma =\textbf{decode}(\gamma',\delta)$. If the decoding is successful, $Charlie$ confirms that he got the correct recovering result.

\section{Anonymous Secret Sharing in a Noisy Quantum Network}

Equipped with the security tools from the previous section, here we analyze the security and performance of our QASS protocol in a noisy quantum network. We consider a noise model in which each qubit is subjected to the same individual noisy channel, which can also encompass noise on the local measurements performed on the state\cite{ref30}. To ensure the anonymity of the receiver, the recovery process needs the assistance of the non-secret receiver, so the possibility of the non-secret receiver being controlled by an adversary must be considered.

\subsection{Security in the Presence of Noise}

In the noise model mentioned above, suppose each qubit is individually affected by a noise map $\Lambda$ while being transmitted to the nodes. if $|W\rangle \langle W|_{n+1}$ is the $(n+1)$-particle W state prepared by $Alice$, then after transmitting,
\begin{align}
\label{eqn.3}
\omega _{n+1}^\Lambda = \Lambda ^{\otimes n+1}(|W\rangle \langle W|_{n+1})
\end{align}
is the actual state distributed to the parties at step (3) of protocol \hyperref[Protocol 4]{4}. In what follows we will show that our protocol is perfectly secure in the active adversary scenario in the noisy network defined by the above equation.

According to the definition of the Permutational-invariance preserving map\cite{ref21}, the noise channel of our interest, $\Lambda ^{\otimes n+1}$, preserves permutational invariance due to the tensor structure. Accordingly, we will prove the following Theorem:

\textbf{Theorem 6}\textbf{.} Our QASS protocol is receiver-anonymous in the active adversary situation in the noisy quantum network modeled by $\omega _{n+1}^\Lambda$.

\textit{Proof.} Following the definition of a permutational-invariance-preserving map, the noise channel introduced in our protocol, preserves permutational invariance due to the tensor structure. So the proof of Theorem 5 follows the same steps as the proof of Theorem 4, the difference is that the state $|W\rangle \langle W|_{n+1}$ is replaced by $\omega _{n+1}^\Lambda$. The guessing probability of a receiver $Bobr_i$ is given by
\begin{align}
&P_{guess}[br_i|\mathcal{W}^{\mathcal{D}}, \mathcal{C}, br_i\notin \mathcal{D}]\\
= & \mathop{max}\limits_{M^j} \sum\limits_{b_j\in \mathcal{H}} P[br_i = b_j|br_i\notin \mathcal{D}]Tr[M^j \cdot \rho_{\mathcal{W}^{\mathcal{D}}, \mathcal{C}|br_i = b_j}^{\Lambda}]\nonumber \\
\leqslant & \mathop{max}\limits_{b_j\in \mathcal{H}} P[br_i = b_j |br_i\notin \mathcal{D}].\nonumber
\end{align}

Therefore, $P_{guess}$ satisfies our anonymity requirement in Definition 2, and Theorem 6 is proved.

In a realistic quantum network, it is impossible to ensure that all qubits are subjected to the action of the same noise channel. So we would like to analyze in the sense that each qubit experiences a slightly different noise, following Definition 3. Then the total state of the noisy channels is given by
\begin{align}
\label{eqn.4}
\hat{\omega}_{n+1}^{\hat{\Lambda}} = \bigotimes\limits_{i=1}^{n+1} \Lambda_i (|W\rangle \langle W|_{n+1}),
\end{align}
where $|| \Lambda - \Lambda_i||_1 \leq \varepsilon_i$, $||\cdot||_1$ represents the 1-norm of a matrix, $\varepsilon_i \to 0$, which is the parameter in $\varepsilon$-receiver anonymity.

\textbf{Theorem 7}\textbf{.} Our QASS protocol is $\varepsilon$-receiver-anonymous in the active dishonest participant situation in the noisy quantum network modeled by $\hat{\omega}_{n+1}^{\hat{\Lambda}}$.

\textit{Proof.} Follows the same steps as the proof of Theorem 4 and Theorem 5, we can calculate
\begin{align}
&P_{guess}[br_i|\mathcal{W}^{\mathcal{D}}, \mathcal{C}, br_i\notin \mathcal{D}]\\
= & \mathop{max}\limits_{M^j} \sum\limits_{b_j\in \mathcal{H}} P[br_i = b_j|br_i\notin \mathcal{D}]Tr[M^j \cdot \hat{\rho}_{\mathcal{W}^{\mathcal{D}}, \mathcal{C}|br_i = b_j}^{\hat{\Lambda}}]\nonumber \\
\leqslant & \mathop{max}\limits_{b_j\in \mathcal{H}} P[br_i = b_j|br_i\notin \mathcal{D}] + (n+1) \varepsilon_{max}, \nonumber
\end{align}
where $\hat{\rho}_{\mathcal{W}^{\mathcal{D}}, \mathcal{C}|br_i=b_j}^{\hat{\Lambda}}$ is the state of the adversaries at the end of the protocol, $\varepsilon_{max}=\mathop{max}\limits_{i\in [\mathcal{H}+\mathcal{R}_\mathcal{H}]} \varepsilon_i$, $(n+1) \varepsilon_{max} \to \varepsilon$. Therefore, $P_{guess}$ satisfies our $\varepsilon$-anonymity requirement in Definition 3, Theorem 5 is proved.

\subsection{Performance in a Noisy Network}

In this section, we analyze the performance of protocol 4 in a noisy quantum network. To complete this task reliably, we assume that all the participants follow the protocol honestly. After step (3), the resultant anonymous entangled state between $Alice$ and $[\mathcal{R}]$ is given by
\begin{align}
    \omega _{m+1}=& \frac{1}{\mathcal{N}} Tr_{n-m} [\Lambda ^{\otimes n+1}(|W\rangle \langle W|_{n+1}) \nonumber \\
    &\cdot (I_{m+1} \otimes |\Vec{0}\rangle \langle \Vec{0}|_{n-m})],
\end{align}
where $|W\rangle \langle W|_{n+1}$ is the $(n+1)$-particle W state shared in advance, $|\Vec{0}\rangle \langle \Vec{0}|_{n-m}$ is a projection onto the $|0\rangle$ state of $(n-m)$ parties and $\mathcal{N}$ is a normalization factor that can be calculated as
\begin{align}
\mathcal{N}=& Tr_{n+1} [\Lambda ^{\otimes n+1}(|W\rangle \langle W|_{n+1}) \nonumber \\ 
&\cdot (I_{m+1} \otimes |\Vec{0}\rangle \langle \Vec{0}|_{n-m})].
\end{align}

In the noiseless case, $Alice$ and $[\mathcal{R}]$ can obtain a perfect W state anonymously. However, the states shared in the noisy channel may deviate from what is expected. Next, we will discuss the performance of the anonymous secret sharing protocol over two types of noisy channels:

1. $\Lambda$ is the dephasing channel, which is modeled by
\begin{align}
\Lambda (\rho)= q\rho + (1-q)Z \rho Z,
\end{align}
where $\rho$ is a single qubit state, $Z$ is the Pauli $Z$ gate, and $q\in [0, 1]$ is the noise parameter.

2. $\Lambda$ is the depolarizing channel, which is modeled by
\begin{align}
\Lambda (\rho)= q\rho + (1-q)\frac{I}{2},
\end{align}
where $\rho$ is a single qubit state, $\frac{I}{2}$ is a maximally mixed state
in two-dimensional Hilbert space, and $q\in [0, 1]$ is the noise parameter.

To confirm the performance of our protocols, we fix the figure of merit to be the fidelity of the obtained anonymous entangled (AE) state with the ideal state that is obtained in the protocol when no noise is present,
\begin{align}
F_{AE} (\omega _{m+1})= Tr[\omega _{m+1} \cdot |W\rangle \langle W|_{m+1}]
\end{align}
where $\omega _{m+1}$ is the anonymous entangled states between $Alice$ and $[\mathcal{R}]$ arising from measuring W states subjected to the network noise, and $|W\rangle \langle W|_{m+1}$ is the $(m+1)$-particle perfect W state.

In what follows we explain what it means for an anonymous entangled state to be useful. According to Ref. \cite{ref31}, not all states are entangled enough to be a resource for teleportation. Besides, the quality of a low-fidelity anonymous entanglement could be further improved by performing entanglement distillation\cite{ref32}. However, entanglement distillation protocols for W states et al. can be carried out only when fidelities of initial states are larger than $\frac{1}{2}$. So we can extend the definition of what it means to say that a resource state is useful for anonymous transmission to multi-particle entangled states. We say that the anonymous entangled state is a useful resource for quantum teleportation if its fidelity is larger than $\frac{1}{2}$, i.e. $F_{AE} > \frac{1}{2}$.

To evaluate the behavior of the protocols, we calculate the fidelity of anonymous entanglement as a function of the noise parameter $q$, the number of participants $n$, and the number of secret receivers $m$, for the depolarizing and dephasing channels. 

1. Dephasing channels.
\begin{align}
F_{AE} (\omega _{m+1})= 2q^2-2q+1
\end{align}

2. Depolarizing channels.
\begin{small}
    \begin{align}
&F_{AE} (\omega _{m+1})= \frac{\Delta_1}{\Delta_2}\\
&\Delta_1= (q+1)\{n(q-1)^2-m(n+3)(q-1)^2 \nonumber \\
&+m^3(6q^2+2)+m^2[(q-1)^2-2n(3q^2+1)]\} \\
&\Delta_2= 4\{n+m(n+3)(q-1)-nq \nonumber \\
&+2m^3(q+1)-m^2[q-1+2n(q+1)]\}
\end{align}
\end{small}

\begin{figure}[h]
		\centering
		\includegraphics[width=1.0\linewidth]{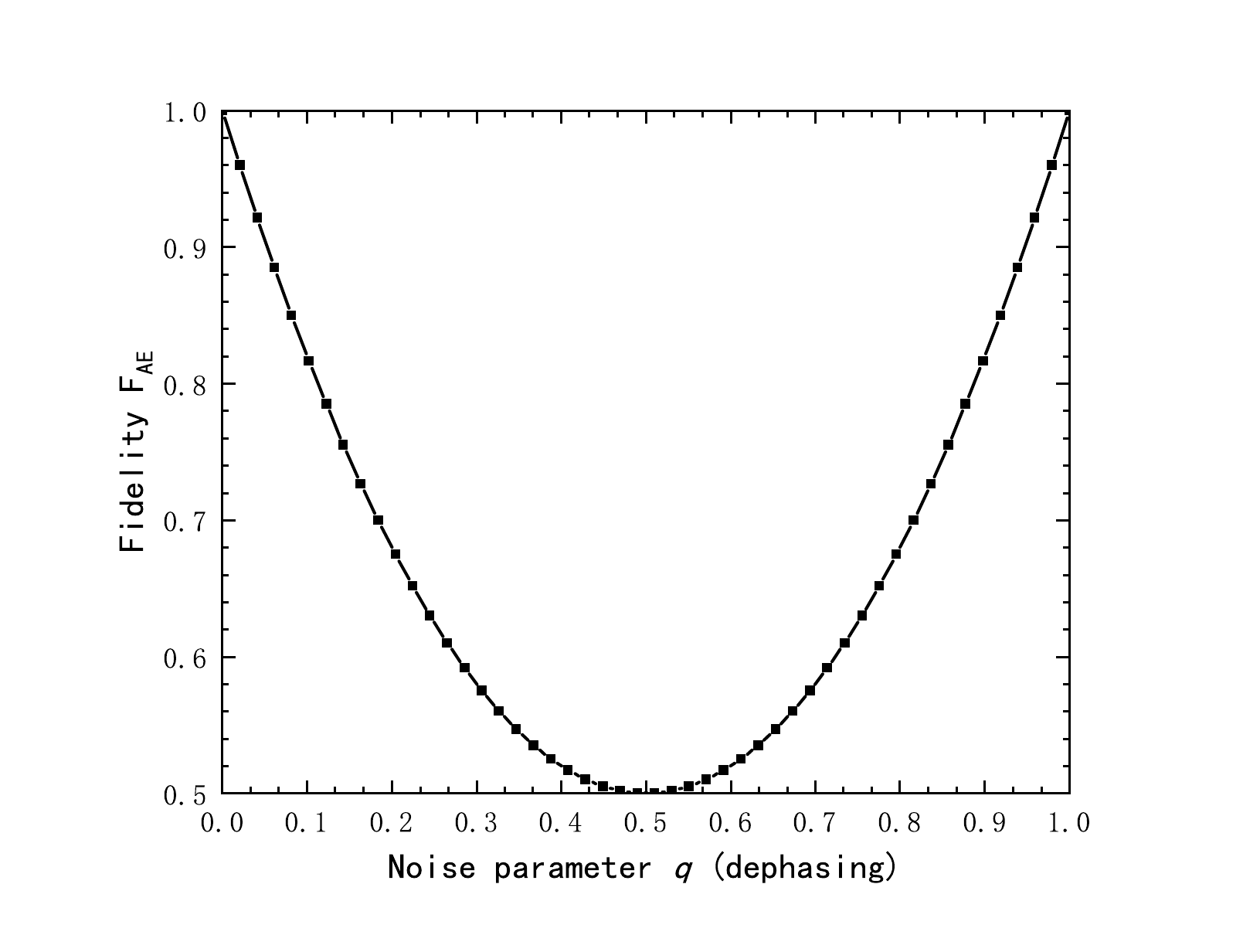}
		\caption{Fidelity of anonymous entanglement as a function of the noise parameter $q$ for dephasing
        noise.}
		\label{FIG:4}
\end{figure}

We start by looking at the dephasing noise. Observe that in this case, the fidelity of anonymous entanglement created with the W state $F_{AE} (\omega _{m+1})$ is irrelevant with $n$ and $m$. Specifically, this implies that when fixed dephasing noise is present in the network, the quality of the anonymous link only depends on the noise parameter, regardless of the number of participants or secret receivers. This results in great performance when there are a large number of participants in the system. The performance of our protocol for dephasing noise is shown in Fig. \ref{FIG:4}.

\begin{figure}[h]
		\centering
		\includegraphics[width=1.0\linewidth]{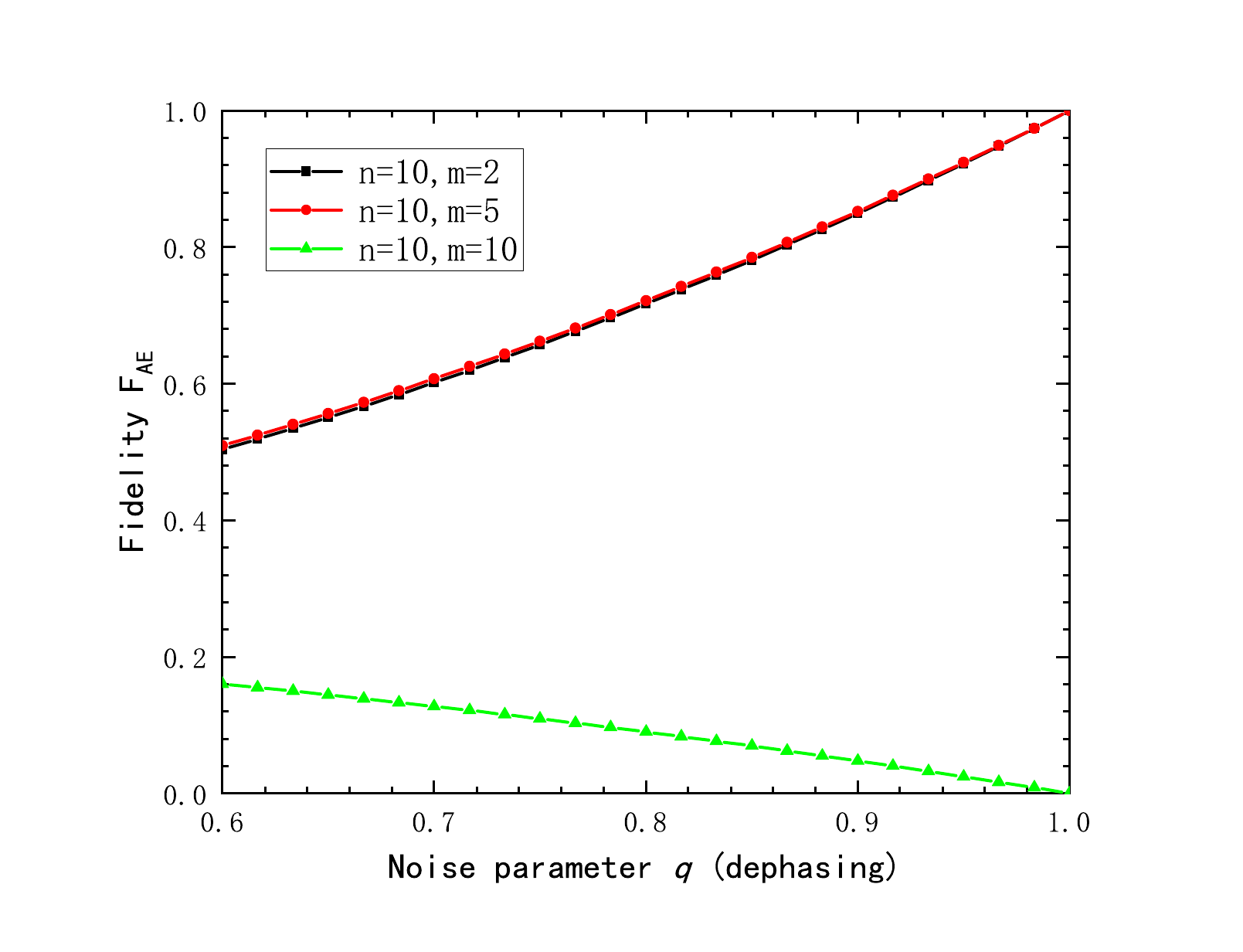}
		\caption{Fidelity of anonymous entanglement as a function of the noise parameter $q$ for depolarizing noise. Examples for $n= 10,m=\{2, 5, 10\}$.}
		\label{FIG:5}
\end{figure}

When depolarizing noise is present in the network, unlike the dephasing noise, the fidelity of the anonymous entanglement generated by our protocol depends on the numbers $m$ and $n$. We first analyze the fidelity affected by $m$ when $n$ is constant. It can be seen in Fig. \ref{FIG:5} that except for the case $n=m$, the fidelity images coincide.

After analyzing more cases of $n$ and $m$, we can know that the fidelity of anonymous entanglement is relatively stable and greater than $\frac{1}{2}$ when $q > \frac{1}{2}$ independent of $m$, except for the case of $n=m$, where the protocol does not meet the requirements of usefulness. In fact, the case $n=m$ corresponds to the case where all participants are secret receivers, which is not common in general.

\begin{figure}[h]
		\centering
		\includegraphics[width=1.0\linewidth]{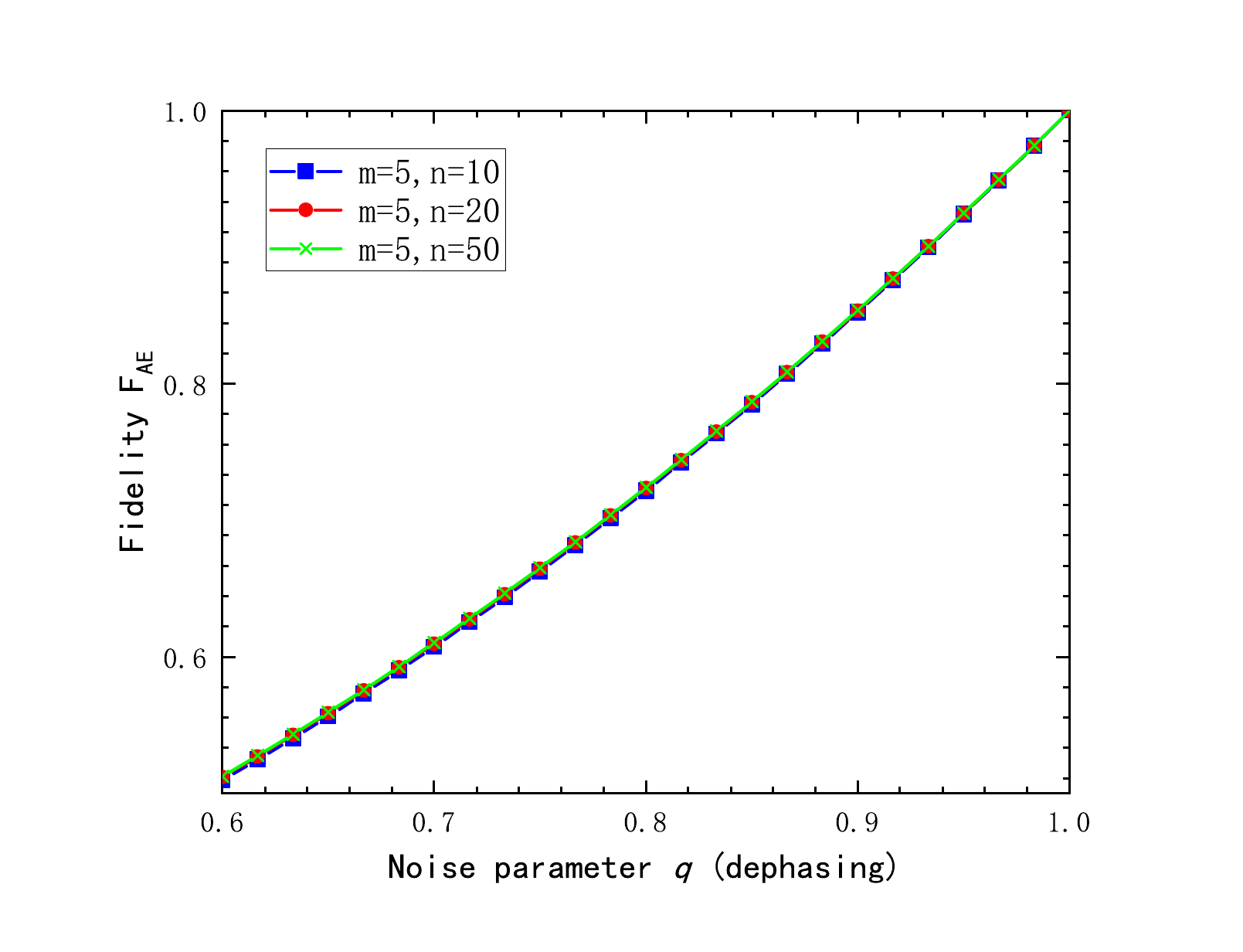}
		\caption{Fidelity of anonymous entanglement as a function of the noise parameter $q$ for depolarizing noise. Examples for $n=\{10, 20, 50\},m=5$.}
		\label{FIG:6}
\end{figure}

Then we analyze the fidelity affected by $n$ when $m$ is constant. It can be seen in Fig. \ref{FIG:6} that the function images coincide even though $n$ is increasing. Thus we can draw a similar conclusion as for dephasing noise, that is, the protocol can still maintain good performance when there are a large number of potential receivers in the system.

\section{Conclusion}
In this research, we have pioneered the integration of quantum mechanics with the realm of quantum secret sharing, culminating in the development of a quantum anonymous secret sharing protocol utilizing W states.   This development represents an essential exploration in quantum information processing, facilitating the secure and anonymous distribution of quantum secrets. It can effectively resist attacks on anonymous receivers and quantum secret information from malicious external adversaries and dishonest authenticated participants. The application of W states within QASS demonstrates substantial efficacy in counteracting noise interference, which is a useful step towards bridging the theoretical constructs of quantum mechanics with their practical implementation in quantum networks. Two interesting future research are the pursuit of more efficient quantum resources to enhance the functionality of quantum anonymous secret sharing, and the exploration of the broader utility of quantum advantages in addressing other practical challenges.

\section{Acknowledgments}

This research was supported by the Key Lab of Information Network Security, Ministry of Public Security (C21605).

\bibliographystyle{plain}

\begin{thebibliography}{40}
\bibitem {ref1} Yu-Guang Yang, Yue-Chao Wang, Yong-Li Yang, Xiu-Bo Chen, Dan Li, Yi-Hua Zhou, and Wei-Min Shi, \href{https://doi.org/10.1007/s11433-021-1692-5}{Science China Physics, Mechanics \& Astronomy \textbf{64(6)}, 260321 (2021).}

\bibitem {ref2} Ying-Hui Yang, Fei Gao, Xia Wu, Su-Juan Qin, Hui-Juan Zuo and Qiao-Yan Wen, \href{https://doi.org/10.1038/srep16967}{Scientific reports \textbf{5(1)}, 16967 (2015).}

\bibitem {ref3} Ignatius W. Primaatmaja, Koon Tong Goh, Ernest Y.-Z. Tan, John T.-F. Khoo, Shouvik Ghorai, and Charles C.-W. Lim, \href{https://doi.org/10.22331/q-2023-03-02-932}{Quantum \textbf{7}, 932 (2023).}

\bibitem {ref4} Cristian Toma, Marius Popa, Catalin Boja, Cristian Ciurea, and Mihai Doinea, \href{https://doi.org/10.3390/electronics11121895}{Electronics \textbf{11(12)}, 1895 (2022).}

\bibitem {ref5} Jae-Geun Song, Sung-Jun Moon, and Ju-Wook Jang, \href{https://doi.org/10.3390/s21123958}{Sensors \textbf{21(12)}, 3958 (2021).}

\bibitem {ref6} Dariush Abbasinezhad-Mood and Morteza Nikooghadam, \href{https://doi.org/10.1109/TIE.2018.2807383}{IEEE Transactions on Industrial Electronics \textbf{65(10)}, 7996--8004 (2018).}

\bibitem {ref7} Debiao He, Huaqun Wang, Muhammad Khurram Khan, and Lina Wang, \href{ https://doi.org/10.1049/iet-com.2016.0091}{IET Communications \textbf{10(14)}, 1795--1802 (2016).}

\bibitem {ref8} Yu-Guang Yang, Bing-Xin Liu, Guang-Bao Xu, Yi-Hua Zhou, and Wei-Min Shi, \href{https://doi.org/10.1109/TIFS.2023.3288989}{IEEE Transactions on Information Forensics and Security \textbf{18}, 4034-4045 (2023).}

\bibitem {ref9} Awais Khan, Uman Khalid, Junaid ur Rehman, and Hyundong Shin, \href{https://doi.org/10.1109/TCOMM.2022.3168079}{IEEE Transactions on Communications \textbf{70(6)}, 4026--4037 (2022).}

\bibitem {ref10} C. Blundo, and D. R. Stinson, \href{https://doi.org/10.1016/S0166-218X(97)89208-6}{Discrete Applied Mathematics \textbf{77(1)}, 13-28 (1997).}

\bibitem {ref11} Guo-Dong Li, Yi-Xi Xu, Qing-Le Wang, Zhi-Hao Zhuang, and Wen-Chuan Cheng, \href{https://doi.org/10.1360/SSPMA-2023-0215}{Scientia Sinica Physica, Mechanica \&amp; Astronomica \textbf{53}, 179-190 (2023).}

\bibitem {ref12} Song Lin, Gong-De Guo, Feng Huang, and Xiao-Fen Liu, \href{https://doi.org/10.1103/PhysRevA.93.012318}{Physical Review A \textbf{93(1)}, 012318 (2016).}

\bibitem {ref13} Wei Huang, Qiao-Yan Wen, Bin Liu, Qi Su, and Fei Gao, \href{https://doi.org/10.1103/PhysRevA.89.032325}{Physical Review A \textbf{89(3)}, 032325 (2014).}

\bibitem {ref14} Yue-Ran Li, Dong-Huan Jiang, and Xiang-Qian Liang, \href{https://doi.org/10.1007/s11128-021-03288-6}{Quantum Information Processing \textbf{20}, 1--33 (2021).}

\bibitem {ref15} Weiyang Ke, Run-hua Shi, Hui Yu, and Xiaotong Xu, \href{https://doi.org/10.1088/1402-4896/acd3bd}{Physica Scripta \textbf{98(9)}, 095116 (2023).}

\bibitem {ref16} Xinchao Ruan, Hang Zhang, Yiyu Mao, Zhipeng Wang, Zhiyue Zuo, and Ying Guo, \href{https://doi.org/10.1364/OE.471000}{Optics Express \textbf{30(23)}, 41204--41218 (2022).}

\bibitem {ref17} Lang Jiang, Guangqiang He, Ding Nie, Jin Xiong, and Guihua Zeng, \href{https://doi.org/10.1103/PhysRevA.85.042309}{Physical Review A \textbf{85(4)}, 042309 (2012).}

\bibitem {ref18} Wei Yang, Liusheng Huang, and Fang Song, \href{https://doi.org/10.1038/srep26762}{Scientific reports \textbf{6(1)}, 26762 (2016).}

\bibitem {ref19} Christopher Thalacker, Frederik Hahn, Jarn de Jong, Anna Pappa and Stefanie Barz, \href{https://doi.org/10.1088/1367-2630/ac1808}{New Journal of Physics \textbf{23(8)}, 083026 (2021).}

\bibitem {ref20} Anupama Unnikrishnan, Ian J. MacFarlane, Richard Yi, Eleni Diamanti, Damian Markham, and Iordanis Kerenidis, \href{https://doi.org/10.1103/PhysRevLett.122.240501}{Physical review letters \textbf{122(24)}, 240501 (2019).}

\bibitem {ref21} Victoria Lipinska, Gláucia Murta, and Stephanie Wehner, \href{https://doi.org/10.1103/PhysRevA.98.052320}{Physical Review A \textbf{98(5)}, 052320 (2018).}

\bibitem {ref22} Zhan-jun Zhang, Yong Li, and Zhong-xiao Man, \href{https://doi.org/10.1103/PhysRevA.71.044301}{Physical Review A \textbf{71(4)}, 044301 (2005).}

\bibitem {ref23} Wei Huang, Qiao-Yan Wen, Bin Liu, and Fei Gao, \href{https://doi.org/10.1088/1674-1056/24/7/070308}{Chinese Physics B \textbf{24(7)}, 070308 (2015).}

\bibitem {ref24} Dmitri Horoshko and Sergei Kilin, \href{https://doi.org/10.1016/j.physleta.2011.01.038}{Physics Letters A \textbf{375(8)}, 1172-1175 (2011).}

\bibitem {ref25} Frederik Hahn, Jarn de Jong, and Anna Pappa, \href{https://doi.org/10.1103/PRXQuantum.1.020325}{PRX Quantum \textbf{1(2)}, 020325 (2020).}

\bibitem {ref26} Lv-zhou Li and Dao-wen Qiu, \href{https://doi.org/10.1088/1751-8113/40/35/010}{Journal of Physics A: Mathematical and Theoretical \textbf{40(35)}, 10871 (2007).}

\bibitem {ref27} Yu-Guang Yang, Xiao-Xiao Liu, Shang Gao, Yi-Hua Zhou, Wei-Min Shi, Jian Li, and Dan Li, \href{https://doi.org/10.1103/PhysRevA.104.052415}{Physical Review A \textbf{104(5)}, 052415 (2021).}

\bibitem {ref28} Chang ho Hong, Jino Heo, Jin Gak Jang, and Daesung Kwon, \href{https://doi.org/10.1007/s11128-017-1681-0}{Quantum Information Processing \textbf{16}, 1--20 (2017).}

\bibitem {ref29} Jaewoo Joo, Young-Jai Park, Sangchul Oh, and Jaewan Kim, \href{https://doi.org/10.1088/1367-2630/5/1/136}{New Journal of Physics \textbf{5(1)}, 136 (2003).}

\bibitem {ref30} Zeng-Rong Zhou, Yu-Bo Sheng, Peng-Hao Niu, Liu-Guo Yin, Gui-Lu Long, and Lajos Hanzo, \href{https://doi.org/10.1007/s11433-019-1450-8}{Science China Physics, Mechanics \& Astronomy \textbf{63(3)}, 230362 (2020).}

\bibitem {ref31} Michał Horodecki, Paweł Horodecki, and Ryszard Horodecki, \href{https://doi.org/10.1103/PhysRevA.60.1888}{Physical Review A \textbf{60(3)}, 1888 (1999).}

\bibitem {ref32} Yu-Bo Sheng, Jun Pan, Rui Guo, Lan Zhou, and Lei Wang, \href{https://doi.org/10.1007/s11433-015-5672-9}{Science China Physics, Mechanics \& Astronomy \textbf{58}, 1-11 (2015).}

\end{thebibliography}

\end{document}